# Particle image velocimetry and modelling of horizontal coherent liquid jets impinging on and draining down a vertical wall

Aouad, W., Landel, J.R., Dalziel, S.B., Davidson, J.F. and Wilson, D.I.*
Department of Chemical Engineering and Biotechnology, New Museums Site, Pembroke Street, Cambridge, CB2 3RA, UK
Department of Applied Mathematics and Theoretical Physics, Centre for Mathematical Sciences, Wilberforce Road, Cambridge CB3 0WA, UK

*Corresponding author

Dr D. Ian Wilson
Department of Chemical Engineering and Biotechnology
New Museums Site
Pembroke Street
Cambridge
CB2 3RA, UK

*E-mail* diw11@cam.ac.uk
*Tel* +44 1223 334791
*FAX* +44 1223 334796

# Particle image velocimetry (PIV) and modelling of horizontal coherent liquid jets impinging on and draining down a vertical wall

Aouad, W., Landel, J.R., Dalziel, S.B., Davidson, J.F. and Wilson, D.I.*
Department of Chemical Engineering and Biotechnology, New Museums Site, Pembroke Street, Cambridge, CB2 3RA, UK
Department of Applied Mathematics and Theoretical Physics, Centre for Mathematical Sciences, Wilberforce Road, Cambridge CB3 0WA, UK

**Abstract**

The flow patterns created by a coherent horizontal liquid jet impinging on a vertical wall at moderate flow rates (jet flowrates 0.5-4.0 L min$^{-1}$, jet velocities 2.6-21 m s$^{-1}$) are studied with water on glass, polypropylene and polymethylmethacrylate (acrylic, Perspex®) using a novel particle image velicometry (PIV) technique employing nearly opaque fluid doped with artificial pearlescence to track surface velocity. Flow patterns similar to those reported in previous studies are observed on each substrate: their dimensions differed owing to the influence of wall material on contact angle. The dimensions are compared with models for (*i*) the radial flow zone, reported by Wang *et al*. (2013b), and the part of the draining film below the jet impingement point where it narrows to a node. For (*ii*), the model presented by Mertens *et al.* (2005) is revised to include a simpler assumed draining film shape and an alternative boundary condition accounting for surface tension effects acting at the film edge. This refined model gives equally good or better fits to the experimental data. The effective contact angle which gives good agreement with the data is found to lie between the measured quasi-static advancing and receding contact angles, at approximately half the advancing value. The PIV measurements confirmed the existence of a thin fast moving film with radial flow surrounding the point of impingement, and a wide draining film bounded by ropes of liquid below the impingement point. While these measurements generally support the predictions of existing models, these models assume that the flow is steady. In contrast, surface waves were evident in both regions and this partly explains the difference between the measured surface velocity and the values estimated from the models.

*Highlights (suggestions)*

- Substrate nature affects the dimensions of the wetted regions and flow behaviour
- Surface velocity in the radial flow zone and falling film measured using PIV technique
- Liquid spreading upwards forms ropes of liquid bounding the falling film
- Dimensions of wetted region on 3 substrates predicted reasonably well by models
- Narrowing of falling film region modelled by modified Mertens *et al*. (2005) model
- Surface waves of different types are evident.

## 1. Introduction

Impinging liquid jets are widely used in cleaning operations for removing soiling layers on process vessels (Jenssen *et al*., 2011), walls (Birch *et al*., 2008) and in dishwashers (Pérez-Mohedano *et al*., 2015). The liquid is usually water or an aqueous solution of surfactants and other detergent species. When a coherent jet impinges on a surface, liquid flows away from the point of impingement in a radial pattern until a point where the fast moving thin film changes to a deeper, slower moving state. When a vertical jet impinges downwards on a horizontal plate this gives rise to a circular hydraulic jump, which has been studied extensively since the initial work of Watson (1964) (see, for example, Craik *et al*., 1981; Bohr *et al*., 1993; Bush *et al*., 2006). When the jet impinges at an angle, the hydraulic jump is elliptical and this has been modelled successfully by Kate *et al*. (2007) and Blyth and Pozrikidis (2005).

When a jet impinges on a vertical wall the flow pattern is no longer cylindrically axisymmetric, owing to gravity, and the liquid falls downwards to give a range of wetting patterns determined by the flow rate, the fluid properties and the surface-liquid interaction manifested in the contact angle, β. Knowledge of this wetting behaviour is important for cleaning operations involving walls and inclined surfaces, as the removal of soiling layers or contaminants is determined by a combination of shear stress, material transport, and soaking (time spent in contact with cleaning solution) (Wilson, 2005).

This paper follows on from a series of studies of the flow behaviour and cleaning performance associated with liquid jets impinging on vertical and inclined walls (Wilson *et al*., 2012; Wang *et al*., 2013a, 2013b; Wilson *et al*., 2014; Wang *et al*., 2015). The series builds on a model presented in Wilson *et al*. (2012) which uses a relatively simple momentum balance to describe the geometry and velocity field of the radial flow zone surrounding the impingement point (see below). In the present study, our aim is to study experimentally the geometry and, where possible, the velocity field of the different regions of the flow produced by a horizontal coherent jet impinging on a vertical surface.

This is the first time that detailed experimental measurements of the velocity field of a thin falling film produced by an impinging jet are presented. The novel particle imaging velocimetry technique of Landel *et al.* (2015) can capture, at high time and space resolutions, the two-dimensional velocity field at the surface of a thin film that is not constrained in a channel. In our experiments, it is essential that the film remains unconstrained at the edges, but rather flows freely on a planar surface. Indeed, the force due to surface tension effects, acting at the edges of the film, is key to understand the physics of the flow produced by an impinging jets. As modelled by Wilson *et al.* (2012), the size of the radial flow zone is controlled primarily by surface tension acting at the film boundary. In the model of Mertens *et al.* (2005), the narrowing and subsequent braiding pattern observed in the draining flow downstream of the radial flow zone is also controlled by surface tension forces acting at the edges.

## 1.1 Anatomy of the impingement pattern

Figure 1 shows photographs and a schematic identifying the characteristic features of a horizontal jet impinging at point O on an otherwise dry vertical wall. The terminology which follows is that employed in our previous studies (Wilson *et al.*, 2012; Wang *et al.*, 2013a). Short videos of the impingement region and the draining film are provided as supplementary material V1 and V2, respectively.

AA is the horizontal line passing through O. At and above AA, the liquid flows radially outwards from O until the change in depth, which we term the *film jump* in order to distinguish it from the hydraulic jump on horizontal surfaces. This *radial flow zone* (RFZ) is labelled Zone I and has radial dimension $R$ along AA (where $\theta = 90°$), and extends to height $R_0$ directly above O. Beyond this radius $R$, the liquid flows circumferentially downwards in a rope, in Zone II, with outer radial dimension $R_c$ along plane AA (Figure 1(*b*)). The flow in the rope is typically much thicker than flow in the RFZ. Symmetry suggests that the volumetric flow rate in each rope crossing AA between $R$ and $R_c$, if there is no splashing (*i.e.* loss as droplets of spray), is close to $Q/4$, where $Q$ is the flow rate in the jet. As the flow in the RFZ will be influenced by gravity, the flow will be slightly less than Q/4. Above AA we employ cylindrical co-ordinates based on point O.

Below AA, liquid still flows radially away from O within the RFZ but the film jump is less pronounced and is not evident in much of this region. In these experiments the additional sideways momentum provided by the remaining liquid from the jet causes the wetted region to expand until it reaches a maximum at plane BB, of width $W$. The zone between the horizontal lines AA and BB is labelled Zone III. The photographs (Figure 1(*b,c*)) indicate that ropes of similar width still exist in this region and there are many surface waves forming a circular pattern. Wang *et al.* (2013a) found that the vertical extent of Zone III (from AA to BB) is greater than or equal to $R$. In Zone III the flow out from the impingement point still provides a source of horizontal momentum, acting against the surface tension trying to narrow the film. The net result is a slight increase in the width of the film. Below Zone III there is no longer significant addition of horizontal momentum and surface tension causes the film to narrow, giving Zone IV. (If low flow rates are used, surface tension will cause the wetted region to contract below AA, see Wilson *et al.* (2012)). Between AA and BB we employ Cartesian co-ordinates; $z$ is the distance downstream from AA.

Zone IV is marked by a number of features: the continuation of the rope on each boundary, of significant width and with a thickness much larger than in the interior of the flow (see darker edges in Figure 1(*c*)); the interior region bounded by the ropes in which almost horizontal crests of surface waves of varying wavelength move downwards, which we term the *film region*; and *nodes* where the two ropes meet and create a knot of liquid which then spreads out again further downstream. On a tall plate, multiple generations of nodes can be observed (see Mertens *et al.*, 2005), and, under certain flow conditions, the flow can form several streams (termed 'braiding' by Mertens *et al.*, 2004). Cartesian co-ordinates are again used in Zone IV: $x$ is the distance downstream from plane BB; $y$ is the lateral co-ordinate, directed to the right; the local half-width of the film is $w$.

### 1.2 Structure

The objective of this study is to analyse experimentally the different flow regions produced by a steady horizontal coherent jet impinging on a vertical flat plate. We compare predictions of different models, including a model based on that of Mertens *et al.* (2005) proposed in this paper, with our experimental measurements for the geometry and the velocity field of the flow. The experimental investigation of the velocity field of the thin film flow is made possible for the first time thanks to the novel technique of Landel *et al.* (2015).

The models are described in section 2. In section 3 we present our experimental procedure and apparatus. We also describe briefly the velocity measurement technique of Landel *et al.* (2015). In section 4, we compare the predictions of the models of Wilson *et al.* (2012), Wang *et al.* (2013b), Mertens *et al.* (2005) and our model which is based on that of Mertens *et al.* (2005) with our experimental data for the four different flow zones described in Figure 1. The predictions and hypotheses of these different models are discussed. We draw conclusions in section 5.

## 2 Modelling

The following models describe the flow patterns arising when a horizontal jet impinges normally against a vertical wall. Models for Zone I where the jet does not impinge normally are reported in Wang *et al*. (2013b, 2015).

### 2.1. Zone I, predicting $R$

In the absence of gravity/body forces, Wilson *et al*. (2012) modelled the radial flow outwards from the point of impingement in the radial flow zone as a thin film with a parabolic velocity profile, *i.e*. a laminar Nusselt film. The mean radial velocity, $U$, at radius $r$, is given by

$$\frac{1}{U} = \frac{1}{U_o} + \frac{10\pi^2 \nu}{3Q^2}\left[r^3 - r_o^3\right], \qquad [1]$$

where $\nu$ is the liquid kinematic viscosity; $r_0$ and $U_0$ are the radius and mean velocity of the (cylindrical) jet, respectively. Wang *et al*. (2013b) modified this to include gravity, giving

$$\frac{d}{dr}\left(\frac{QU}{2\pi}\right) = -5\pi\frac{\nu U^2 r^2}{Q} - \frac{5}{6}\frac{Qg\cos\theta}{2\pi U}, \qquad [2]$$

where $g$ is the acceleration due to gravity and $\theta$ is the angle at which the streamline is inclined to the vertical. For horizontal flow along AA, $\theta = \pm 90°$ and the solution of [2] is simply [1].

This model does not account for the distance required to develop the parabolic velocity profile in the thin film, which was determined for the laminar case by Watson (1964). This distance scales with $r_0$ and is of order 2-3$r_0$ for the tests reported here. Equations [1] and [2] are therefore not expected to be accurate for low flow rates, when the RFZ is small. The model also assumes steady flow in the expanding film, which precludes the formation of surface waves that are evident in the photograph (Figure 1(*b*)). The assumption of the parabolic velocity profile will be investigated briefly in Section 4.1 by comparing the surface velocity, $U_s$, measured by the PIV technique with that predicted by the models: for a parabolic film, $U_s = 3U/2$.

In the above models, the location of the film jump, at $r = R$ along AA, Figure 1(*a*) - and at $r = R_0$ when vertically above O - is determined by a momentum balance in which the momentum flux associated with the flow outwards is balanced by the surface tension force acting inwards, as shown in Figure 2, viz.

$$\frac{6}{5}\rho U_R^2 h_R = \gamma\left(1 - cos\,\beta\right) \quad [3]$$

Substituting this result into [1], and assuming that $r_0^3$ and $1/U_0$ are small, yields:

$$R = \left[\frac{9\rho^2 Q^3}{50\pi^3 \mu\gamma\,(1-\cos\beta)}\right]^{1/4} = 0.276\left[\frac{\rho^2 Q^3}{\mu\gamma\,(1-\cos\beta)}\right]^{1/4} \quad [4]$$

A similar result is obtained for $R_0$, but the presence of the gravitational term requires numerical integration of equation [2] in order to obtain $U(r)$ to substitute into equation [3].

## 2.2. Zone II, estimating $R_c$, and Zone III

The rope region extends beyond $R_0$ to reach the maximum height, $R_z$, directly above the impingement point (see Figure 1(*a*)). Wang *et al.* (2013b) presented a simple model for the width of the rope region (Zone II), which varies with angular coordinate $\theta$. In this model, the flow rate of liquid in the rope is proportional to $\theta$. The model is derived from a momentum balance on an element of rope, assuming negligible wall shear. A detailed description is given in Wang *et al.* (2013b). The rope is assumed to have a circular cross-section of diameter $D$, given by

$$D(\theta) = \frac{2}{\pi}\sqrt{\frac{Q}{\sqrt{2gR}}}\sqrt{\frac{\theta}{\sqrt{f_{rope}(\theta)}}}, \quad [5]$$

where $\theta$ is in radians and

$$f_{rope}(\theta) = \frac{2(\cos\theta - 1)}{\theta^2} + \frac{2\sin\theta}{\theta} - \cos\theta \quad [6]$$

As $\rightarrow 0$, $f_{rope} = \theta^2/4 - \theta^4/36$ and the asymptotic limit of [5] is

$$D_0 = \frac{2}{\pi}\sqrt{Q\sqrt{\frac{2}{gR}}} \quad [7]$$

$R_z$ is then calculated from:

$$R_z = R_0 + D_0 \quad [8]$$

The flow in the rope is complex. At higher flow rates the rope edge is often unsteady so that the rigid substrate is subject to random wetting and dewetting. Contact angles measured under quiescent conditions, whether retreating or advancing, are unlikely to give a full description of the contact line. Wang *et al.* (2013b) presented a first order model for the width of the rope which underestimated the measured values. The complexity of the physics at the contact line renders a full computational simulation, including the free surface, a demanding task. In the absence of a reliable predictive model, Wang *et al.* demonstrated that the width of the wetted region at the AA plane, $R_c$ in Figure 1(*a*), was related to $R$ by the empirical relationship $4/3 \leq R_c/R \leq 2$, the constant of proportionality varying with substrate nature (and contact angle) and flow rate.

Zone III is similarly complex and the finding that the length of this region is approximately equal to $R$ has been stated above. Wang *et al.* (2013a) found that the maximum width, $W$, depended on the substrate nature and presence of surfactants, which is expected as these determine the wetting behaviour. Their study also showed that surfactants had little effect on $R$, as the time taken for the liquid to transit the RFZ was short and there was insufficient time for mass transfer of surfactant to reach its equilibrium concentration on the freshly formed surface. A quantitative model for Zone III has yet to be developed.

2.3 Zone IV

The model for the evolution of the width of a finite stream of liquid flowing down an inclined plane presented by Mertens *et al.* (2005) is adapted here to describe the narrowing of the film in Zone IV. Symmetry allows the flow to be modelled in terms of $Q/2$ in each half of the domain as shown in Figure 3: $x$ is the distance downstream from line BB and the local half-width of the film is $w$. Mertens *et al.* postulated that the height of the film, $h(y)$, could be described by a quartic of the form

$$h = \left(w^2 - y^2\right)\left(a - by^2\right), \qquad [9]$$

Parameters $a$ and $b$ can be evaluated by relating the gradient at the edge to that given by the contact angle, which yields

$$a = \frac{\tan\beta}{2w} + bw^2, \qquad [10]$$

Several assumptions about the flow in the film (see Mertens *et al.*, 2005) allow $b$ to be obtained from the flow rate, $Q$, via

$$b = \frac{15}{8w^5}\left(\frac{Q}{2u} - \frac{w^2 \tan\beta}{3}\right), \quad [11]$$

in which $u$ is the local average velocity in the falling film. Mertens *et al.* combined the equations of motion and continuity to give a pair of coupled ordinary differential equations. In dimensionless form these are:

$$\frac{d}{dx_*}\left(u_* \frac{dw_*}{dx_*}\right) = F_* - \Pi_1 u_*^2 w_*^2 \frac{dw_*}{dx_*}, \quad [12]$$

$$u_* \frac{dw_*}{dx_*} = 1 - \Pi_2 u_*^3 w_*^2. \quad [13]$$

The dimensionless variables are

Velocity $$u_* = u\frac{2\gamma}{\rho Q g}, \quad [14]$$

Length ($l = x, y$ or $h$) $$l_* = l\frac{4\gamma^2}{\rho^2 Q^2 g}. \quad [15]$$

For a jet of water at 20°C with flow rate 1 L $min^{-1}$, representative of the experiments reported here, the characteristic scaling velocity is 1.04 m $s^{-1}$ and the length scale is 0.11 m. The constants $\Pi_1$ and $\Pi_2$ are defined

$$\Pi_1 = 3\frac{Q}{2}\frac{\mu\rho g}{\gamma^2}, \quad [16]$$

$$\Pi_2 = 3\left(\frac{Q}{2}\right)^5 \frac{\mu\rho^6 g^4}{\gamma^7}. \quad [17]$$

In dimensionless form, Mertens *et al.*'s narrowing force $F^*$ in Equation [12] is given by

$$F^* = \frac{1}{16}\left(15\frac{\Pi_1}{\Pi_2}\frac{1}{u_* w_*^2} - 5\tan\beta\right)\left(15\frac{\Pi_1}{\Pi_2}\frac{1}{u_* w_*^2} - \tan\beta\right). \quad [18]$$

The pair of ODEs [12-13] has initial conditions: $u_*$(0), which is estimated using a fraction of the jet velocity, $U_o$, as $u$(0); $w_*$(0), based on $W$/2, taken from experiments; and $du_*/dx_*$ (0), which is set at zero along BB.

Equations [12] and [13], with $F^*$ given by Equation [18], are referred to here as the *Mertens Model*. The equations were integrated numerically using Matlab™ on a desk-top PC. Initial calculations established that the flow pattern and narrowing behaviour were relatively insensitive to $u(0)$, and a value of $u(0) = U_o/5$ was used in the calculations presented here. The effect of other values of $u(0)$ is discussed in Section 3.3.

Mertens *et al.* reported very good agreement between their experimental results and the above model. However, inspection of the quartic relationship in Equation [9] reveals that it can predict negative film heights, which is physically infeasible. Indeed, tests reproducing the results reported in their paper (Figure 3) indicated regions of negative film thickness. For the experimental conditions employed in the tests reported here, Equation [9] gave negative film thicknesses in the central part of the film for much of the region upstream of the node.

The Mertens film profile was therefore discarded and the flow modelled as the simplest case, *i.e.* a Nusselt falling film of uniform height, $h(x)$, except at the edges where the gradient was given by the contact angle (solid line in Figure 3(*b*)). This is in effect the simplest limiting case, as it ignores the presence of the ropes and, as such, does not require any assumptions about their shape or the velocity profile within them. It cannot predict the existence of the nodes and braiding, which is one of the strengths of the Mertens *et al.* model. These shortcomings of the simpler film profile are not important for cleaning applications, which motivated this study.

This uniform film model depends upon the alternative boundary condition

$$F_* = \cos\beta - 1 \quad . \qquad [19]$$

Equations [12] and [13], with $F^*$ given by Equation [19], is termed the *Revised Model*. This model only applies until the first node is reached. Other, explicit, models for the rope could be introduced but would require calibration and would not change the essence of the physics in the domain of interest.

*Contact angles*

At BB (see Figure 3) the nature of the contact line switches. In Zones II and III, above BB, surface tension acts to stop the liquid expanding outwards, while below BB, in Zone IV, surface tension acts to narrow the flow. To a first approximation, the advancing contact angle, $\beta_{advancing}$, might be expected to describe the behaviour above BB, and the receding contact angle, $\beta_{receding}$, expected to be relevant below BB. Both these parameters, however, are normally measured using quasi-static tests (as described in the Experimental section). In the present films the contact line is not moving, but the fluid moves parallel to it. The contact line takes up a steady position, but the dynamic pressure forces on the fluid side are unsteady (due to turbulent fluctuations or simply periodic waves). This is investigated here by employing an *effective contact angle*, which is obtained by fitting the data to the model predictions for Zones I and IV. The effective contact angle is compared with the measured values of $\beta_{advancing}$ and $\beta_{receding}$, and lessons drawn.

## 3. Experimental

### 3.1 Apparatus

Liquid, either softened tap water or a dyed solution in the case of the PIV tests, was pumped from a 20 L holding tank through a rotameter and a 500 mm long tube of internal diameter 4 mm before entering the nozzle. The tube, with length of 125 diameters, gave a well-developed flow upstream of the nozzle. Two stainless steel nozzles, with inner convergent angle 45° and orifice diameters of 2 mm and 3 mm, were used. The pressure upstream of the entry section was measured to monitor the flow rate. Both pressure and rotameter readings were calibrated in separate catch-and-weigh tests. The distance from the nozzle to the target plate was set at 5 cm in order to ensure that the jet was coherent. The angle of inclination of the substrate was checked regularly using an electronic inclinometer to maintain a vertical plane (within ±0.2°). The flow rates employed in tests reported here varied from 0.48 L $min^{-1}$ to 4.0 L $min^{-1}$ at 20 °C, corresponding to jet Reynolds numbers, defined $Re_{jet} = U_o r_o/\nu$, of 2,600 to 21,200 for the 2 mm diameter nozzle.

The target plates were 300 mm wide × 300 mm long × 5 mm thick sheets of borosilicate glass, polypropylene or Perspex® (polymethylmethacrylate) mounted on an aluminium frame. The plates were colourless so that the flow patterns could be photographed from behind as well as in front. Liquid draining from the plate was recycled back to the feed tank. Still photographs were taken with a Canon Digital IXUS 75 camera. The surface roughness of the plates was measured using a contact profilometer which gave average roughness (Ra) values of approximately 0.008 μm for the glass and 0.02 μm for the Perspex and polypropylene sheets. Between tests the target was cleaned with soap and distilled water, followed by an isopropanol wash.

### 3.2 Particle image velocimetry

PIV measurements employed solutions of 6.5 g $L^{-1}$ methylene blue dye (Fisher Scientific) in softened tap water with 0.17 volume % artificial pearlescence (Iriodin 120 pigment, Merck,) added as tracer particles. These are flat, titanium dioxide coated mica particles with size 5-25 μm: they exhibit a silver-pearl colour when mixed with water. The dye concentration was chosen so that the solution was opaque and only particles on or very near the surface of the flow would be imaged, so that the PIV captured surface velocities (see Landel *et al.*, 2015, for more details). The steady shear rheology of this dilute suspension was measured in a smooth-walled Couette cell on an ARES controlled strain rheometer (Rheometric Scientific) and found to be Newtonian with a viscosity similar to that of water. The shapes of the flow patterns obtained with the test rig for the suspension were similar to those obtained with water under the same conditions.

PIV images were captured using a high-speed greyscale camera (Photron-Fastcam SA1.1) fitted with a 60 mm focal length AF Mirco-Nikkor lens. We fitted a UV/IR blocking filter on the lens because the camera was sensitive to the infrared part of the spectrum, which was not absorbed by the solution of methylene blue dye. The flow was illuminated using two 300 W xenon arc lamps. A lens aperture of f/4.0D and shutter speed of 1/30,000 s provided sufficient resolution without too much light, thus limiting the amount of over-illumination due to the light being reflected on the unsteady surface waves. 6×6 cm (512×512 pixels) images of Zone I were recorded at 20,000 frames $s^{-1}$, while 12×12 cm (1024×1024) images of Zone IV were captured at 5400 frames $s^{-1}$. Images were analysed using the DigiFlow software tool (Dalziel *et al.*, 2007).

### 3.3 Contact angle determination

Sessile drop measurements for determining advancing and receding contact angles were performed on glass, polypropylene and Perspex substrates using a DataPhysics OCA system running OCA 20 software. The results obtained with up to 10 repeat measurements are summarised in Table 1. Large differences between advancing and receding contact angles are evident, indicating significant contact angle hysteresis. The values for Perspex are close to those reported by Zografi and Johnson (1984), who also reported noticeable differences between rough and smooth surfaces. The $\beta_{advancing}$ value for polypropylene is similar to that reported by Chibowski (2007), whereas the $\beta_{receding}$ value of 32° is quite different from his value of 79.7°. The $\beta_{advancing}$ value for glass is similar to that reported by Wang *et al.* (2013a), of 39 ±5°. Measurements of contact angle are sensitive to the presence of surface imperfections (which can be introduced by manufacture) as well as adsorbed material (and thus cleaning regime and surface history), so the above comparisons are offered as a guide. The values measured on the test materials reported in Table 1 are used in the calculations.

## 4. Results and Discussion

4.1 Zones I and II: flow patterns and substrates

The effect of flow rate on the size of the radial flow zone, $R$, was studied for different nozzle sizes and substrates. The results are compared with the model predictions (Equation [4]) using an effective contact angle of 90° in Figure 4. There was poor agreement with the model for glass when the measured static advancing contact angle from Table 1 was used to calculate $R$, which Wang *et al.* (2013b) also found for flow rates such as those used here. At high flow rates they found that an effective contact angle of 90° gave good agreement with their observations. Figure 4($a$) shows a similar result. They attributed this to fluctuations in the rope, which was evident in videos of these experiments on glass (see Supplementary Video 1). For Perspex, the measured contact angle of 70° gave reasonable agreement at some flow rates: at others, 90° gave better agreement, and the latter results are plotted in Figure 4(b). The measured contact angle for polypropylene is 90° and Figure 4($c$) again gives good agreement with the model. An effective contact angle of 90° was used in subsequent RFZ calculations.

Figure 5 shows that the Zone I model (Equations [2] and [3]) is able to describe the effect of gravity on the shape of the RFZ. The effect of flow rate on the height of the RFZ above the impingement point, $R_0$, as well as the extent of the rope, $R_z$ (Equation [8]), are predicted reasonably well for all three surfaces. The Perspex $R_z$ data are compared with predictions for $\beta = 70°$ and 90°; the latter gives better agreement. Similarly good agreement for Perspex and glass were reported by Wang *et al.* (2013b): they did not study polypropylene so the current results extend the validity of their model, again with $\beta = 90°$giving the best comparison.

Although the location of the film jump defining the edge of Zone I appears to be best modelled using $\beta = 90°$, independent of the measured static contact angle, the same is not true for Zone II. The width of the rope is strongly affected by the nature of the substrate, with the half-width of the wetted region at the impingement point, $R_c$, tending to be larger on glass, which water wets more readily, than Perspex and polypropylene. When the data were plotted in the form $R_c$ vs. the measured value of $R$, as reported by Wang *et al.* (2013b), the Perspex and polypropylene fitted the relationship $R_c = 4R/3$ reasonably well, while the glass data followed $R_c = 2R$ more closely. When the $R_c$ data were plotted against $4R^*/3$, where $R^*$ is the value predicted by Equation [4] and which includes the contribution from the substrate via the advancing contact angle, Figure 6 shows that all three data sets showed good agreement. The difference for the glass is the effect of rope stability on the measured value of *R*.

The PIV measurements allow us to compute the streamlines of the surface velocity field at a particular instant in time. Figure 7 shows two images of the flow above the impingement point for a flow rate of approximately 0.7 L min$^{-1}$ ($Re_{\mathrm{jet}}$ = 3700) on Perspex (*a*) and on glass (*b*). The presence of surface waves in Zone I is evident, while the instantaneous streamlines (shown with white arrowed solid curves) show the assumption of radial flow to be reasonable. The streamlines tend to curve slightly circumferentially downwards as the film jump merges into the rope. In contrast, flow in the rope (Zone II) is almost completely circumferential. We note that although the streamlines presented in Figure 7 are computed for a particular instant in time and the flow is not truly steady (for instance due to surface capillary waves or velocity fluctuations in the jet), we computed that the streamlines are in average very similar for different times. Therefore, we believe that the flow in the RFZ is in good approximation radial, until it curves slightly circumferentially downwards, close to the film jump, due to gravity.

We have studied qualitatively the waves observed at the surface of the RFZ. Surface waves appear in the RFZ above a certain jet flow rate. For both Perspex and glass, the surface waves were observed for flow rates greater than or equal to 8.2 cm$^3$ s$^{-1}$, corresponding to a jet velocity of 2.6 m s$^{-1}$ (jet Reynolds number 5200). Surface waves were not observed in the RFZ for Perspex or for glass at flow rates less than or equal to 5.7 cm$^3$ s$^{-1}$, corresponding to a jet velocity of 1.8 m s$^{-1}$ (jet Reynolds number 3600). We believe these nonlinear capillary waves are due to a jet instability at the impingement point. It was noticeable that these waves tend to accumulate at the film jump, as revealed in video V1. Some waves are also seen to overtake previous waves before reaching the film jump. It also appears that these waves can perturb the flow downstream of the film jump, thus provoking the characteristic long wave instability (see *e.g.* Kalliadasis *et al.* (2011) for more details). The different stages of the long wave instability could be observed in these experiments as the jet flow rate increased: from the onset of the two-dimensional long waves (for the lowest flow rates, less than or equal to 5.7 cm$^3$ s$^{-1}$, jet velocity 1.8 m s$^{-1}$), to the development of the lateral instability (occurring further downstream or at larger flow rates), and a fully unsteady perturbed surface (for flow rates $\geq$ 8.2 cm$^3$ s$^{-1}$ corresponding to a jet velocity of 2.6 m s$^{-1}$ and jet Reynolds number 5200).

Figure 8 shows examples of the time-averaged surface velocity profiles obtained from PIV measurements, for the experiment on Perspex in Figure 7(*a*). All the velocity profiles presented in this paper, unless otherwise stated, are averaged in time over a duration of approximately 1 s, which is much longer than the longest time scale in the flow (the period of long surface waves). Data are not available for $r < 1$ cm as the camera view is obscured by the nozzle (see Figure 7). Data are presented for the vertical (Figure 8(*a*)) and horizontal (Figure 8(*b*)) directions. Plotted alongside the PIV measurements are the jet velocity, $U_o$, and the estimated surface velocity, $U_s$, calculated from Equation [2], which assumes that there is a parabolic velocity profile in the film in Zone I. The relationship between $U_s$ and the depth-averaged average velocity in the film, $U$, is $U_s = 3U/2$.

The measured values differ noticeably from the model predictions in both cases. In the horizontal direction (Figure 8(*b*)) the measured values are consistently larger, decreasing as $r$ increases in a similar manner to the model. In the vertical direction (Figure 8(*a*)), the measured values decrease more strongly with $r$, and cross the predicted trend as $r$ approaches $R_0$. The measured surface velocity will be affected by the presence of the surface waves evident in Figure 7 so it is not surprising that the values do not agree with the theoretical predictions. Moran *et al*. (2002) also reported that in laminar films, the velocity profile at the troughs or crests of waves could depart from the well-known semi-parabolic profile. Further work is required with the PIV technique in order to identify the contribution from waves.

4.2 Zone III Transition to draining film

Wide draining films which subsequently narrowed, as shown in Figure 1(*b*), were obtained with all three substrates at the flow rates studied. The width of the film was recorded at various positions below the impingement level: the results for zone III, *i.e*. the region extending from the impingement level to the position of maximum film width, are presented in Figure 9; this diagram shows the ratio of $w$ to $R_c$, which is half the wetted width along AA (see Figure 1). The streamwise co-ordinates are scaled against $R$, as $z/R$, following the example of Wang *et al.* (2013a), who found that the film stopped spreading horizontally after the radial flow region below the midplane finished; thus $R$ represents a sensible scaling distance.

On Perspex and polypropylene substrates the film widens only slightly beyond $z/R$ = 1 (indicated with a horizontal dashed line), reaching its widest point at $z/R \sim 1.3$-1.5, see Figure 9(*a*),(*c*). Thereafter the film starts to narrow. On glass, however, where water is more strongly wetting (see contact angle values in Table 1), the film continues to spread outwards until $z/R$ = 3-5, see Figure 9(*b*). (If $R^*$ were used instead, $z/R^*$ would still be larger than 1.5). Similar effects on Zone III behaviour were reported by Wang *et al.* (2013a), where they modified the contact angle on glass and Perspex by the addition of Tween 20, an aqueous surfactant, to make the liquid more strongly wetting.

There is no clearly observable film jump below the point of jet impingement in Figure 1(*b*), apart from the relative accumulation of nonlinear capillary waves, which appear at sufficiently large jet flow rates. This is because gravity accelerates the flow downwards in this region. Figure 1(*b*) features an area just below O where the flow is marked by circular surface waves, suggesting that the liquid is flowing radially away from the point of impingent, in a manner similar to the RFZ (Zone I). Comparisons of the surface velocity extracted from PIV measurements with that estimated using the RFZ model for Zone I, (Equation [2], with $\theta$ = 180°) showed similar features to Figure 8: contributions from surface waves were again dominant (data not reported).

4.3 Zone IV Narrowing draining film

Both flow rate and substrate nature affect the flow behaviour in this region. The sections above have demonstrated how these parameters affect the initial half-width of the draining film, $w_o$, measured at BB: in Zone IV the nature of the substrate determines the magnitude of the surface tension force causing the film to narrow, *i.e.* $F^*$, while the flow rate quantifies the mass to be accelerated and thus the time (and length) scale over which narrowing occurs. The shape of the narrowing film was measured at various locations between the point of maximum width, marked BB on Figure 1(*a*), at which $x$ was set to zero, and the next node in the film or the base of the plate, whichever was reached first. It should be noted that the location of maximum width also depended on the flow rate, moving further down the plate as $Q$ increased. Mertens *et al.* (2004, 2005) used a long plate to be able to capture the distance to the node: as the primary interest of the present study is on cleaning and the area contacted by the liquid in the jet, this is not considered essential here.

*Flow pattern*

Figure 10 shows the effect of mass flow rate on the film flow behaviour on Perspex substrates. The initial width, $w_o = W/2$, and distance to the first node increase with flow rate. For the lower flow rates (Figure 10 (*a*, *b*)) the node is reached before the end of the plate, at about $x = 5w_o$. Plotted alongside the experimental data with different lines are the predictions of the Mertens Model (using Equation [18] as the boundary condition for $F^*$), and the Revised Model (which uses our proposed $F^*$ described in Equation [19]). The results obtained with three values of the contact angle are presented in Figure 1(*a*), namely (*i*) the measured (static) receding contact angle, $\beta_{receding}$ (Table 1, 25°, plotted with light grey lines), (*ii*) the advancing contact angle, $\beta_{advancing}$ (Table 1, 70°, plotted with dark grey lines), and (*iii*) the value which gave the best fit to the model across all the data sets, the effective contact angle, $\beta_{fit}$ (plotted with black lines). For Perspex, $\beta_{fit}$ was found to be 35°, which is half the measured $\beta_{advancing}$ value.

The plots show that the measured receding contact angle gives rise to less rapid narrowing, whereas the fitted value gives a reasonably good description of the flow pattern, and one that would be sufficiently accurate for design purposes.

The finding that the measured static receding contact angle overestimates the wetting on the substrate can be interpreted in terms of dynamic phenomena as noted in section 1.1, particularly at large flow rates where the dynamic pressure forces increase. It is evident that the static contact angle does not give an accurate description of this semi-static flow, where the contact line is static but the liquid flows parallel to it with possible fluctuations in speed. With a stationary contact line, one would expect β to be intermediate. In the presence of macroscopic flow fluctuations, the instantaneous value of β (or more precisely, cosβ) is likely to depend on flow conditions. The empirical relationship, $\beta_{fit} \approx 0.5\ \beta_{advancing}$, was also found to apply to the other surfaces tested, as shown by the results on glass (which water wets more strongly) and polypropylene (which is neutral in terms of wetting) in Figure 11. For all three substrates, we find $\beta_{receding} < \beta_{fit} < \beta_{advancing}$. A rough estimation by eye of the thickness of the rope is approximately $h_{rope} \approx 4$ to 5 mm for the experiments on Perspex and $h_{rope} \approx 2$ to 3 mm for the experiments on glass. Then, using the data in Table 2 to estimate the typical width of the ropes, $w_{rope}$, we can compute a rough estimation of the experimental contact angle at the edge of the rope, such that . For Perspex we find that  to , and for glass  to  for the different experiments.

Although these estimations have a large uncertainty they show that the values of  ( and  for glass and Perspex, respectively) used in Figures 10 to 13 are consistent with the experiments.

*Comparison of PIV with model*

The governing equations in Zone IV (Equations [12, 13]) evaluate $u$, the superficial velocity in the film (i.e. the volumetric flow rate per unit width divided by the film depth). This is also a time-averaged velocity as the analysis assumes steady flow. An estimate of the surface velocity, $u_s$, which is measured by PIV, can be generated by assuming that the velocity distribution in the film is parabolic, giving a theoretical estimate $u_s = 3u/2$. The Revised Model is used in these calculations: the Mertens *et al.* model yields qualitatively similar results. The local thickness of the film, δ, can also be estimated from the Nusselt film model (Nusselt, 1916),

$$u_s = \frac{\rho g \delta^2}{2\mu} \quad [20]$$

Figure 12 compares the shape of the falling films and experimental $u_s$ values measured along the falling film centreline ($y = 0$) near the point of maximum width for two substrates at similar flow rates. The two plots in Figure 12(*b*) show the values estimated from the model prediction of $u(x)$ with contact angle $\beta_{fit}$, for different initial starting velocities: (*i*) $u(0) = 0.2U_o$ (plotted using a solid line) and (*ii*) $u(0) = 0.5U_o$ or $u(0) = 0.14U_o$ (plotted using a dashed line). $U_o$ is estimated from $Q = \pi r_o^2 U_o$. In both cases there is little influence of the value of $u(0)$ after 2 cm, and there is good agreement between the model and the measured values within the range of experimental variation.

With Perspex, the experimental surface velocities in the initial few centimetres are noticeably larger than that predicted by the model and that measured on glass. Inspection of the photographs in Figure 12(*a*) provides a plausible explanation of this difference, in that there are many well-defined surface waves (with correspondingly higher values of $u_s$) on the Perspex. The falling film on the glass is wider, and the periodicity of the surface waves consequently different.

The photographs in Figure 12(*a*) also show that the rope, formed in Zone II, persists in Zone IV. Its width, which can be gauged from the shadow pattern and the measurements in Figure 10, does not change noticeably as the film flows downwards. For all four flow rates in Figure 10, the rope width is approximately 1 cm. The presence of the rope is also evident in the surface velocity distributions presented in Figure 13. The film region is evident near the centreline ($y$ < 1.4 cm on Perspex, Figure 13(*a*); $y$ < 2.5 cm on glass, Figure 13(*b*)) with uniform $u_s$, which also suggests, from Equation [20] that the flow is of uniform depth. At larger lateral distance $y$, $u_s$ increases with $y$ before decreasing to zero at the contact line, consistent with the presence of a thicker rope of curved cross-section. Moreover, in both cases the measured $u_s$ values agree with those predicted by the model in the central part of the film, within the rope. This provides some support for the use of the fitted contact angle as this has been shown to give reasonable estimates of both the film width (an overall measure) and the value of the local uniform velocity, ( is assumed uniform along lateral distance ).

The presence of the ropes in both the photographs and velocity profiles is a reminder that our proposed Revised Model does not capture all the physical phenomena present, as we neglected the thicker ropes in the height profile of the film. The PIV results can be used to estimate the flow in the rope region by assuming that the local volumetric flow rate per unit width, Γ, (also known as the wetting rate) is given by the Nusselt analysis (1916) for a stable, wide falling film:

$$\Gamma = \frac{\rho g \delta^2}{3\mu} = 0.943\left(\frac{\mu}{\rho g}\right)^{1/2} u_s^{3/2} \qquad [21]$$

This equation gives only a very rough estimate of the real flow rate since the flow in the experiments is unsteady and highly perturbed. The flow in each region, *i.e.* the rope and uniform film, can then be estimated by integrating along the lateral direction, $\int \Gamma \mathrm{d}y$. This approximation ignores secondary flows and other phenomena, which present a challenging problem (illustrated by the work by Perazzo and Gratton (2004) on the flow of a stable rivulet down an inclined plane).

Table 2 presents data obtained for different cases, one on glass and a second on Perspex, where the flow rate in the rope, $Q_R$, and film region, $Q_F$, were evaluated and are compared with the total, accurately measured flow, $Q$. There is a noticeable difference between $Q_R + Q_F$ and $Q$, up to 36% error, which indicates the degree of approximation involved in these calculations as well as the likely influence of waves. On Perspex $Q_F \approx Q/2$ while on glass $Q_R \approx Q/2$. Wang *et al.* (2013b) proposed that the flow in the two ropes can be estimated as the fraction of the liquid in the impinging jet that passes upwards after striking the target. They found that this simple geometric model gave a reasonable prediction of the flow rate in the falling film for different jet impingement angles, and the tendency for the falling film to form dry patches. This geometric model suggests the tests in Table 2 should give $Q_R = Q_F$. The data, subject to large uncertainty, show scatter and $Q_R \approx Q_F$. We believe that the experimental velocity measurements could be improved by reducing the over-illumination occurring particularly in the ropes, due to the highly unstable surface waves.

Knowledge of the wetting rate (or mean velocity) allows the Reynolds number in the film region to be evaluated. The film Reynolds numbers for the tests reported in Table 2, estimated from $Re_{\text{film}} = uh/\nu$, all lie in the range 130-170.

More detailed investigation, including methods to determine instantaneous film thickness, are required in order to be able to predict the flow behaviour *a priori*. The value of the Mertens Model is that it allows the complex flow pattern in Zone IV to be written in the form of an ODE which can be solved relatively quickly using standard tools. The Revised Model presented here shows that a simplified flow cross-section (see Figure 3(*b*)) yields a reasonably good description of the observed behaviour. The Revised Model offers a method for estimating wetted areas for cleaning applications, but it does not describe braiding. Both models highlight the inadequacy of static contact angle measurements to describe the contact line forces. Resolving the role of momentum in these quasi-static contact lines, and the nature of the effective contact angle, requires further analysis. Detailed computational simulations employing volume of fluid approaches such as reported by Gunjal *et al.* (2005) could also be employed for these systems.

## 5. Conclusions

The flow patterns created by liquid jets impinging on vertical surfaces were investigated using water and a number of different transparent substrate materials: glass, polypropylene and Perspex. The surface velocities were measured using a novel PIV technique based on artificial pearlescence in water dyed opaque so that only particles located at the flow surface were tracked (Landel *et al.*, 2015). The jet velocities studied were relatively low compared to industrial cleaning jet flows, as this gave stable films and little splashing.

The flow behaviour observed matched that reported by previous workers (Wilson *et al*., 2012; Wang *et al*., 2013a, 2013b; Wilson *et al*., 2014; Wang *et al*., 2015): (*i*) Zone I, Figure 1, the formation of a rapid and thin radial film near the point of impingement; (*ii*) Zone II, above the impingement point where gravity caused the liquid to turn and flow downwards as ropes; (*iii*) below the impingement point, a region labelled Zone III with a central falling film bounded by the ropes which widened until the outward momentum was balanced by the surface tension at the contact line acting inwards, and (*iv*) Zone IV, a narrowing region similar to Zone III which terminated in a node, after which the film widened again.

The PIV studies highlighted the presence of surface waves in Zones I, II, III and IV. Surface velocity measurements confirmed that these velocities changed significantly with radial distance in Zones I and III. The distribution of surface velocities in Zone IV indicated that the falling film between the ropes was almost uniform in thickness, and that about half the flow remained in the ropes, as predicted by Wang *et al*. (2013b).

The dimensions of the different zones were compared with the predictive models reported by Wilson and co-workers (Zones I and II; see above) and Mertens *et al*. (Zone IV). In both cases, better quantitative agreement with the experimental data was obtained when the contact angle employed in the calculations was allowed to vary from the measured contact angle. In Zone I, the diameter of the RFZ could be predicted reliably using the measured advancing contact angle until higher flow rates, when a value of 90° gave good agreement. The difference between this effective contact angle and the measured static angle is attributed to the instability along the static contact line arising from perturbations in the flow of liquid in the rope. In Zone IV, the effective contact angle $\beta_{fit}$ obtained by fitting was approximately half that of the measured advancing contact angle and found such that $\beta_{receding} < \beta_{fit} < \beta_{advancing}$. The surface velocity measured by PIV in Zones I and IV was consistently larger than that estimated from the models. This is attributed to the prevalence of surface waves, which were noticeable in many of the experiments.

Two versions of the Mertens *et al*. model were compared with the experimental data: the original, which features a film depth profile which can give unphysical results, and our proposed one where the film depth profile is assumed to be uniform across most of the width of the falling film; of course the film thickness must fall to zero at each edge. The latter model gives a simplified force boundary condition. The latter, revised model gave equally good or better agreement with the experimental data and is suitable for estimating the shape of the draining film, to the first node, for applications such as cleaning.

## Acknowledgements

The apparatus was constructed by Tao Wang and Lee Pratt. Preparatory work by Huifeng Wu, and Neville Research Fellowship for JRL from Magdalene College, Cambridge, are gratefully acknowledged.

# Nomenclature

## Roman

| | | |
|---|---|---|
| $a$ | Constant in the height function, $h(x, y)$, Equation [9] | $m^{-1}$ |
| $b$ | Constant in the height function, $h(x, y)$, Equation [9] | $m^{-3}$ |
| $D$ | Rope width | m |
| $d_N$ | Nozzle diameter | mm |
| $F$ | Capillary force acting on the half-braid | N $m^{-1}$ |
| $F_*$ | Dimensionless form of the capillary force, $F^* = F/\gamma$ | - |
| $g$ | Gravitational acceleration | m $s^{-2}$ |
| $h$ | Height of the film | m |
| $h_R$ | Height of the film at the film jump | m |
| $l$ | Characteristic length | m |
| | Total flow rate | $m^3$ $s^{-1}$ |
| | Flow rate in the central region of the falling film | $m^3$ $s^{-1}$ |
| | Flow rate in the rope region of the falling film | $m^3$ $s^{-1}$ |
| $R$ | Radius of film jump at mid plane | m |
| $R^*$ | Value of $R$ predicted using Equation [4] | m |
| $R_c$ | Outer radius of flow at mid plane | m |
| $R_0$ | Radius of film jump, vertically above O, $\theta = 0$ | m |
| $R_z$ | Outer radius of film jump, vertically above O, $\theta = 0$ | m |
| $r_o$ | Jet radius | m |
| $r$ | Radial co-ordinate | m |
| | Critical film Reynolds number, defined | - |
| $Re_{jet}$ | Jet Reynolds number, defined $Re_{jet} = U_o r_o/\nu$ | - |
| $Re$ | Falling film Reynolds number | - |
| $U$ | Mean velocity in RFZ film | m $s^{-1}$ |
| $U_o$ | Initial mean velocity in RFZ film | m $s^{-1}$ |
| $U_R$ | Film mean velocity at $R$ | m $s^{-1}$ |
| $U_s$ | Surface velocity in RFZ film | m $s^{-1}$ |
| $u$ | Downwards velocity of the draining film | m $s^{-1}$ |
| | Dimensionless form of $u_x$ | - |
| $u_s$ | Surface velocity of the draining film | m $s^{-1}$ |
| | | |
| $V$ | Characteristic velocity | m $s^{-1}$ |
| $W$ | Maximum film width | m |
| $w$, $w_o$ | Half width, half width at $x = 0$ | m |
| $w_*$ | Dimensionless half width | - |
| $x$ | Distance downstream from plane BB, in Zone IV | m |
| $x_*$ | Dimensionless distance downstream from plane BB | - |
| $y$ | Lateral distance from centreline, in Zone IV | m |
| $z$ | Distance downstream from plane AA, in Zone III | m |

## Acronyms

| | | |
|---|---|---|
| PIV | Particle image velocimetry | - |
| RFZ | Radial flow zone | - |

## Greek

| | | |
|---|---|---|
| | Contact angle | $^o$ |
| $\beta_{advancing}$ | Advancing contact angle | $^o$ |
| $\beta_{fit}$ | Contact angle derived from data fitting, Figure 11 | $^o$ |
| $\beta_{receding}$ | Contact angle | $^o$ |
| | Thickness of the falling film | m |
| | Angle from vertical | $^o$ |
| | Surface tension | N $m^{-1}$ |
| $\Gamma$ | Wetting rate | $m^2$ $s^{-1}$ |
| | Dimensionless group in Equation [12], defined in [16] | - |
| | Dimensionless group in Equation [13] , defined in [17] | - |
| $\mu$ | Dynamic viscosity | Pa s |
| $\nu$ | Kinematic viscosity | $m^2$ $s^{-1}$ |
| | Liquid density | kg $m^{-3}$ |

## References

Birch, W., Carré, A. and Mittal, K.L (2008) *Developments in Surface Contamination and Cleaning*. William Andrew, Inc, pp. 693-723.

Blyth, M.G. and Pozrikidis, G. (2005) Stagnation-point flow against a liquid film on a plane wall, *Acta Mechanica,* 180, pp. 203-219.

Bohr, T., Dimon P. and Putkaradze, V. (1993) Shallow water approach to the circular hydraulic jump. *Journal of Fluid Mechanics*, vol. 254, pp. 635-648.

Bush, J.W.M., Aristoff, J.M. and Hosoi, A.E. (2006) An experimental investigation of the stability of the circular hydraulic jump. *Journal of Fluid Mechanics*, vol. 558, pp. 33-52.

Chang, H-C. (1994) Wave evolution on a falling film. *Annual Review of Fluid Mechanics,* vol. 26, pp. 103-136.

Chibowski, E. (2007) On some relations between advancing, receding and Young's contact angles. *Advances in Colloid and Interface Science*, vol. 133, pp. 51–59.

Craik, A.D.D., Latham, R.C., Fawkes, M.J. and Gribbon, P.W.F. (1981) The circular hydraulic jump. *Journal of Fluid Mechanics*, vol. 112, pp. 347-362.

Dalziel, S.B., Carr, M., Sveen, J.K. and Davies, P.A. (2007) Simultaneous synthetic Schlieren and PIV measurements for internal solitary waves. *Measurement Science and Technology,* vol. 18, pp. 533–547.

Davis, M.L. (2010) *Water and wastewater engineering: Design principles and practice.* McGraw-Hill, Dubuque.

Dietze, G.F., Leefken, A. and Kneer, R. (2008) Investigation of the backflow phenomenon in falling liquid films. *Journal of Fluid Mechanics*, vol. 595, pp. 435–459.

Gunjal, P.R., Ranade, V.V. and Chaudri, R.V. (2005) Dynamics of drop impact on solid surface experiments and VOF simulations, *AIChEJ*, 51, pp. 59-78.

Jeffery, G. B. (1922) The Motion of Ellipsoidal Particles Immersed in a Viscous Fluid. *Proceeding of the Royal Society of London. Series A vol.* 102**,** pp. 161–179.

Jensen, B.B.B. (2014) Industrial cleaning flows – Knowledge gap in tank cleaning. *Fouling & Cleaning in Food Processing 2014*, Cambridge, United Kingdom, p. 298.

Jensen, B.B.B., Nielsen, J.B., Falster-Hansen, H. and Lindholm, K-A. (2011) Tank cleaning technology: Innovative application to improve clean-in-place (CIP). Yearbook 2011-12, European Hygienic Engineering and Design Group (EHEDG), pp.26-30, Frankfurt.

Kachanov, Y. (1994) Physical Mechanisms of Laminar Boundary-Layer Transition. *Annual Review of Fluid Mechanics vol.* 26**,** pp. 411–482.

Kate, R. P., Das, P. K. and Chakraborty, S. (2007) Hydraulic jumps due to oblique impingement of circular liquid jets on a flat horizontal surface. *J. Fluid Mech*., vol. 573, pp. 247-263.

Kalliadasis, S., Ruyer-Quil, C., Scheid, B. and Velarde M.G. (2011) *Falling Liquid Films.* Springer.

Landel, J.R., McEvoy, H. and Dalziel, S.B. (2015) Cleaning of viscous drops on a flat inclined surface using gravity-driven film flows. *Food and Bioproducts Processing*, vol. 93, pp. 310-317.

Liu, J., Schneider, J. B. and Gollub, J. P. (1995) Three-dimensional instabilities of film flows. *Physics of Fluids* vol. 7**,** pp. 55-67.

Mertens, K., Putkaradze, V. and Vorobieff, P. (2004) Braiding patterns on an inclined plane. *Nature*, vol. 430, p.165.

Mertens, K., Putkaradze, V. and Vorobieff, P. (2005) Morphology of a stream flowing down an inclined plane. Part 1. Braiding, *Journal of Fluid Mechanics*, vol. 531 (1), pp. 49–58.

Moran, K., Inumaru, J. and Kawaji, M. (2002) Instantaneous hydrodynamics of a laminar wavy liquid film, *International Journal of Multiphase flow*, vol. 28 (1), pp. 731–755.

Nusselt, W. (1916) Oberflachen kondensation des wasserdampfes. *Z. Ver. Dtsch. Ing.*, vol. 60, pp. 541–546 (and pp. 569–575).

Perazzo, C.A. and Gratton, J. (2004) Navier-Stokes solutions for parallel flow in rivulets on an inclined plane, *J. Fluid Mech*., vol. 507, pp. 367-379.

Pérez-Mohedano, R. Letzelter, N., Amador, C., Van der Roest, C.T. and Bakalis, S. (2015) Positron Emission Particle Tracking (PEPT) for the analysis of water motion in a domestic dishwasher, *Chemical Engineering Journal*, vol. 259, pp. 724-736.

Savas, O. (1985) On flow visualization using reflective flakes. *Journal of Fluid Mechanics* vol. 152**,** pp. 253–248.

Trefethen, L. N., Trefethen, A. E., Reddy, S. C. and Driscoll, T. A. (1993) Hydrodynamic stability without eigenvalues. *Science* 261**,** pp. 578–584.

Wang, T., Davidson, J.F. and Wilson, D.I. (2013a) Effect of surfactant on flow patterns and draining films created by a static horizontal liquid jet impinging on a vertical surface at low flow rates. *Chemical Engineering Science*, vol. 88, pp. 79-94.

Wang, T., Faria, D., Stevens, L.J., Tan, J.S.C., Davidson, J.F. and Wilson, D.I. (2013b) Flow patterns and draining films created by horizontal and inclined coherent water jets impinging on vertical walls. *Chemical Engineering Science*, vol. 102, pp. 585–601.

Wang, T., Davidson, J.F. and Wilson, D.I. (2015) Flow patterns and cleaning behaviour of horizontal liquid jets impinging on angled walls, *Food and Bioproducts Processing*, Vol. 93, pp. 333-342.

Watson, E.J. (1964) The radial spread of a liquid jet over a horizontal plane. *Journal of Fluid Mechanics*, vol. 20 (3), pp. 481–499.

Wilson, D.I. (2005) Challenges in cleaning: recent developments and future prospects. *Heat Transfer Engineering*, vol. 26 (1), pp. 51–59.

Wilson, D.I., Le, B.L., Dao, H.D.A., Lai, K.Y., Morison, K.R. and Davidson, J.F. (2012) Surface flow and drainage films created by horizontal impinging liquid jets. *Chemical Engineering Science*, vol. 68, pp. 449–460.

Wilson, D.I., Köhler, H., Cai, L., Majschak, J-P. and Davidson, J.F. (2015) 'Cleaning of a model food soil from horizontal plates by a moving vertical water jet', *Chem. Eng. Sci.* vol. 123, pp. 450-459.

Yeckel, A. & Middleman, S. (1987) Removal of a viscous film from a rigid plane surface by an impinging liquid jet. *Chem. Eng. Commun.* vol. 50**,** pp. 165–175.

Zografi, G. and Johnson, B. A. (1984) Effects of surface roughness on advancing and receding contact angles. *International Journal of Pharmaceutics*, vol. 22, pp. 159-176.

## Tables

Table 1 Measured contact angles

| Material | $\beta_{advancing}$ | $\beta_{receding}$ | Hysteresis $\beta_{advancing} - \beta_{receding}$ |
|---|---|---|---|
| Glass | 36.13.1° | 11.66.4° | ~ 25° |
| Perspex | 70.71.9° | 25.25.6° | ~ 45° |
| Polypropylene | 90.2 3.6° | 32.35.0° | ~ 58° |

Table 2 Partitioning of flow between rope and falling film in Zone IV

| Substrate | $Q$ | $x$ | $W(x)$ | Width of film region /cm | $Q_F/Q$ | $Q_R/Q$ | $(Q_R + Q_F)/Q$ |
|---|---|---|---|---|---|---|---|
| | /L min$^{-1}$ | /cm | /cm | | /% | /% | /% |
| Perspex | 0.48 | 2.29 | 4.44 | 2.1 | 42-47 | 69 | 111-116 |
| | 0.63 | 3.23 | 5.87 | 3.4 | 52-58 | 78 | 130-136 |
| | 0.72 | 4.10 | 5.69 | 3.4 | 45-51 | 35 | 80-86 |
| Glass | 0.56 | 5.94 | 7.60 | 4.6 | 70 | 53 | 123 |
| | 0.72 | 3.40 | 9.76 | 6.0 | 71 | 56 | 127 |

**List of Figure captions**

Figure 1 Flow pattern generated by a horizontal liquid jet impinging normally on a vertical wall: (*a*) Schematic, and photographs from PIV testing of (*b*) region around impingement point (shadow of jet and nozzle visible), and (*c*) draining film and node. Zones and dimensions are described in the text. Experimental conditions: water on Perspex, 20°C, $Q$ = 0.5 L min$^{-1}$ ( = 8.33 g s$^{-1}$), O is the point of jet impingement. Dashed line in (a) represents the film jump.

Figure 2 Schematic of a slice through the flow in Zones I and II showing elements in the film jump.

Figure 3. Geometry and dimensions of Zone IV, below BB. (*a*) Coordinates: CC is an arbitrary plane distance $x$ downstream of the widest point (BB, Figure 1), where the width is $W$. (*b*) Schematic of film thickness profiles. The dashed line is a sketch of the Mertens *et al*. quartic function (Equation [9]). The solid line represents the simple thin film flow advocated in this work, where the flow is assumed to fall as a Nusselt film. Not to scale.

Figure 4 Comparison of predicted film jump radius from Equation [4] with $\beta$ = 90° for (*a*) glass, (*b*) Perspex, and (*c*) polypropylene surfaces. Legend indicates flow rate and nozzle size. $d_N$ = 4 mm indicates no nozzle (open pipe).

Figure 5 Comparison of measured and predicted extent of the radial flow zone directly above the point of impingement for (*a*) glass, (*b*) Perspex, and (*c*) polypropylene using an effective contact angle of 90°.  for Perspex with $\beta$ = 70° also shown. The nozzle diameter, $d_N$, is 3 mm.  The predicted values of $R_0$ were obtained from Equations [2] and [3]; $R_z$ was estimated using Equation [8].

Figure 6 Measured half-width of wetted region on AA, $R_c$, plotted against $R_c = 4/3R^*$, where $R^*$ is predicted using Equation [4] for (*a*) glass; (*b*) Perspex; and (*c*) polypropylene. $d_N$ = 4 mm indicates no nozzle (open pipe).

Figure 7 PIV images of the upper part of the RFZ for (*a*) Perspex, $Q$ = 0.73 L min$^{-1}$, and (*b*) glass, $Q$ = 0.71 L min$^{-1}$. $d_N$ = 2 mm. Nozzle obscures point of impingement O. Dashed lines show boundary of rope (Zone II). White loci with arrows show instantaneous streamlines identified by PIV.

Figure 8 Comparison between the predicted surface velocity and measured values for the flow pattern in Figure 7(a), $Q$ = 0.73 L min$^{-1}$ on Perspex, with a 2 mm nozzle: (*a*) vertical direction; (*b*) horizontal direction.  Error bars on PIV measurements show experimental uncertainty for selected positions.  Dashed horizontal line shows jet velocity, $U_o$.

Figure 9 Shape of draining film in zone III for water on (a) Perspex; (b) glass; (c) polypropylene. 3 mm nozzle, flow rate shown in legend. Vertical co-ordinate $z$ presented in dimensionless form, $z/R$, where $R$ is the value measured in experiments.

Figure 10 Comparison of wetted region half width, $w$, including rope, in Zone IV on Perspex, $d_N$ = 3 mm. The four diagrams (*a*)-(*d*) show the effect of flow rate: (*a*) $Q$ = 2.3 L min$^{-1}$; (*b*) $Q$ = 1.5 L min$^{-1}$; (*c*) $Q$ = 2.0 L min$^{-1}$, and (*d*) $Q$ = 4.0 L min$^{-1}$. The data points show measurements of film half width (•) and width to inside of rope (o). Loci predicted by the (*i*) Mertens Model and (*ii*) Revised Model, with $\beta_{fit}$ = 35° shown on (*a*-*d*): (*a*) also shows the predictions for $\beta = \beta_{receding}$ = 25° and $\beta = \beta_{advancing}$ = 70°.

Figure 11 Width of the draining film in Zone IV for (*a*) glass, (*i*) $Q$ = 0.8 L min$^{-1}$; (*ii*) $Q$ = 1.5 L min$^{-1}$; and (*b*) polypropylene, (*i*) $Q$ = 2.3 L min$^{-1}$, (*ii*) $Q$ = 2.1 L min$^{-1}$. The nozzle diameter is $d_N$ = 3 mm. Also plotted are the loci predicted by the Mertens and Revised models for the contact angle $\beta_{fit}$.

Figure 12 Comparison of zone IV behaviour: (*a*) photographs and (*b*) surface velocity, $u_s$, at centreline ($y$ = 0) for (*i*) Perspex, $Q$ = 0.63 L min$^{-1}$, and (*ii*) glass, $Q$ = 0.56 L min$^{-1}$. Plots in (*b*) show predictions for the Revised Model with different initial starting velocity, $u(x = 0)$ and effective contact angle, β, of (*i*) 35° and (*ii*) 18°.

Figure 13 Comparison of measured downward surface velocity distribution, $u_s$, with value predicted by the Revised Model in Zone IV (horizontal locus) for (*a*) Perspex, $Q$ = 0.63 L min$^{-1}$, $x$ = 3.23 cm, β = 35°; (*b*) glass, $Q$ = 0.56 L min$^{-1}$, $x$ = 5.93 cm, .

**Supplementary material**

Video V1

Video V2

(*a*)

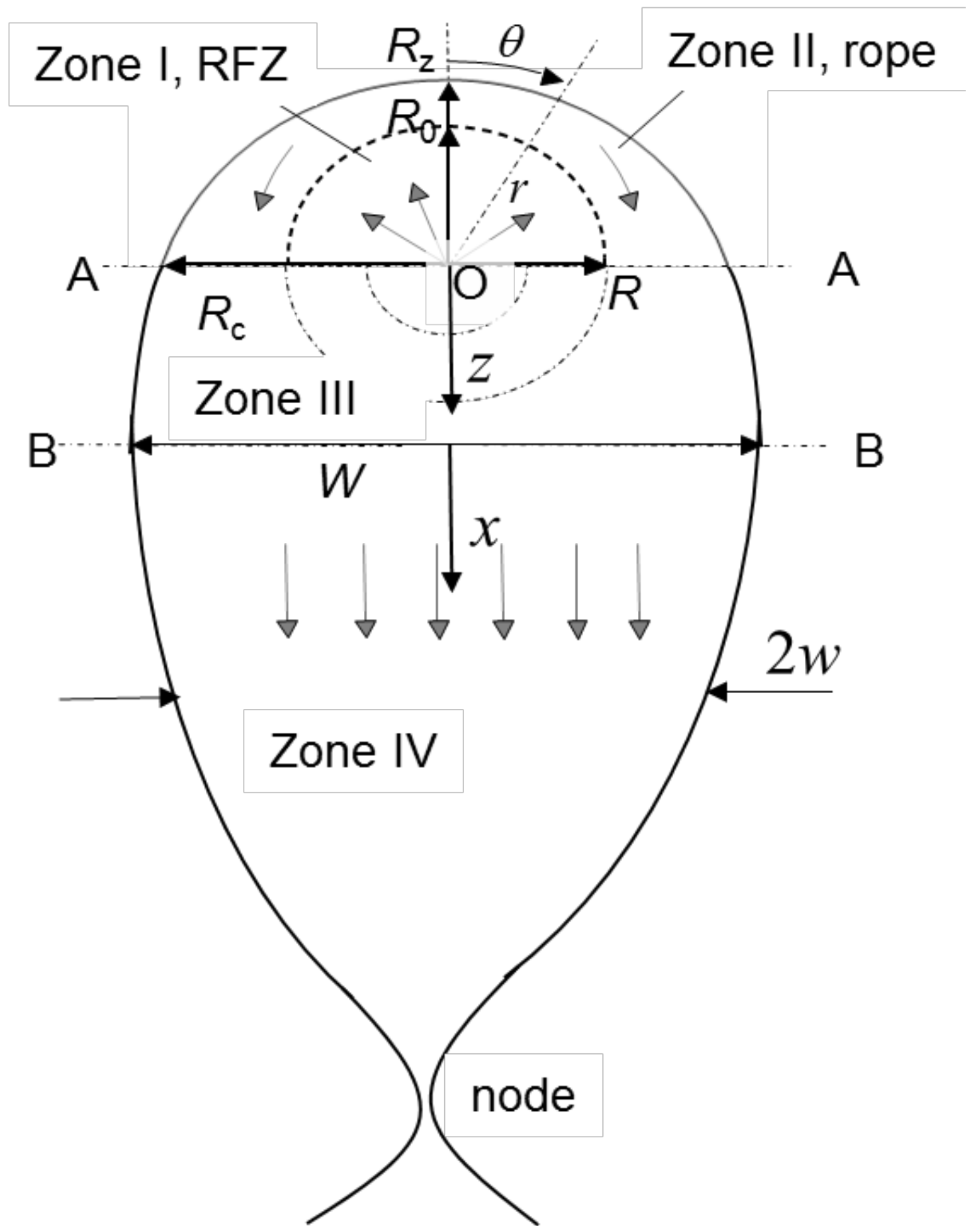

Figure 1 Flow pattern generated by a horizontal liquid jet impinging normally on a vertical wall: (*a*) Schematic, and photographs from PIV testing of (*b*) region around impingement point (shadow of jet and nozzle visible), and (*c*) draining film and node. Zones and dimensions are described in the text. Experimental conditions: water on Perspex, 20°C, $Q$ = 0.5 L min$^{-1}$ ( = 8.33 g s$^{-1}$), O is the point of jet impingement. Dashed line in (*a*) represents the film jump.

(*b*)

(*c*)

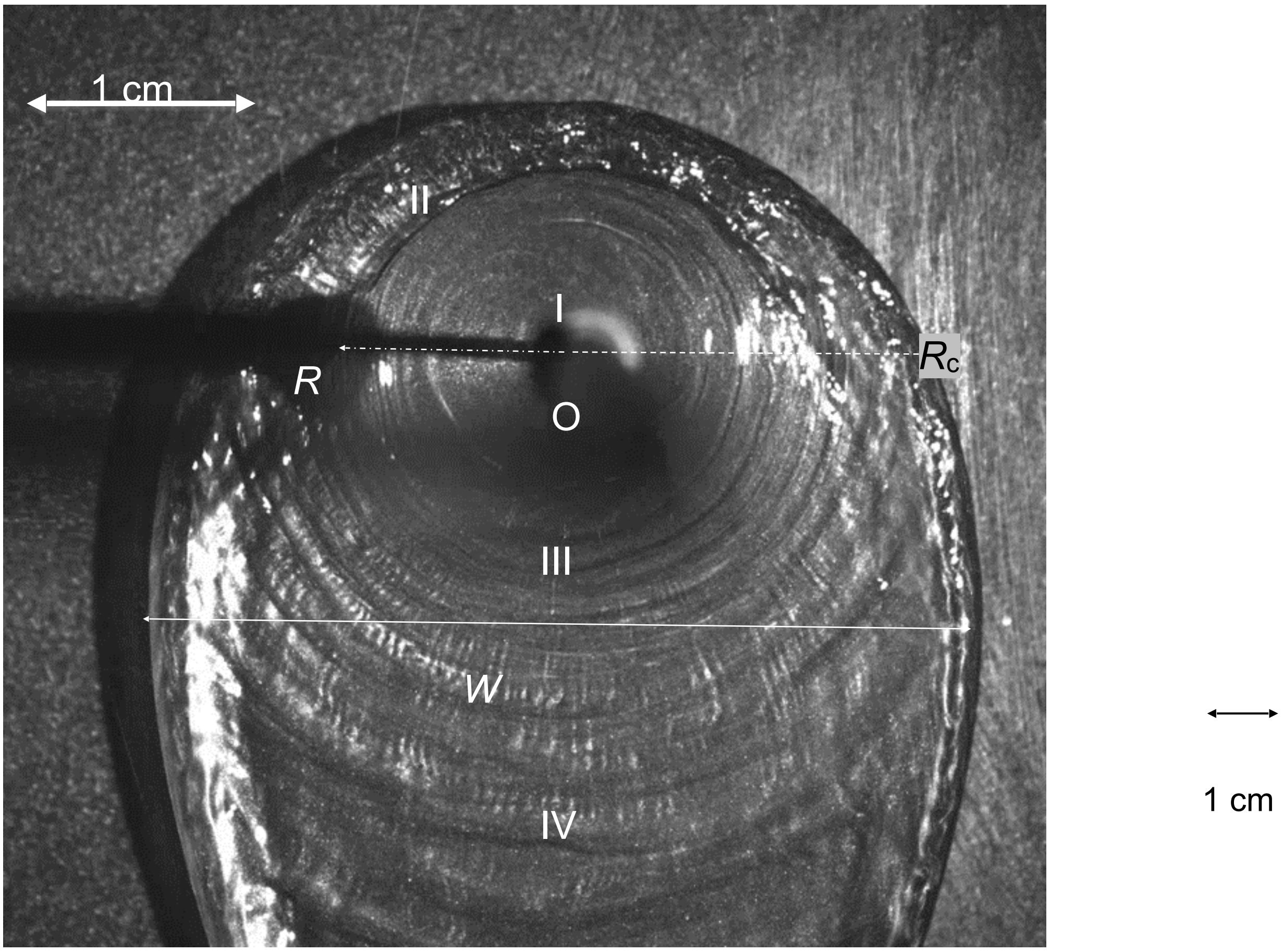

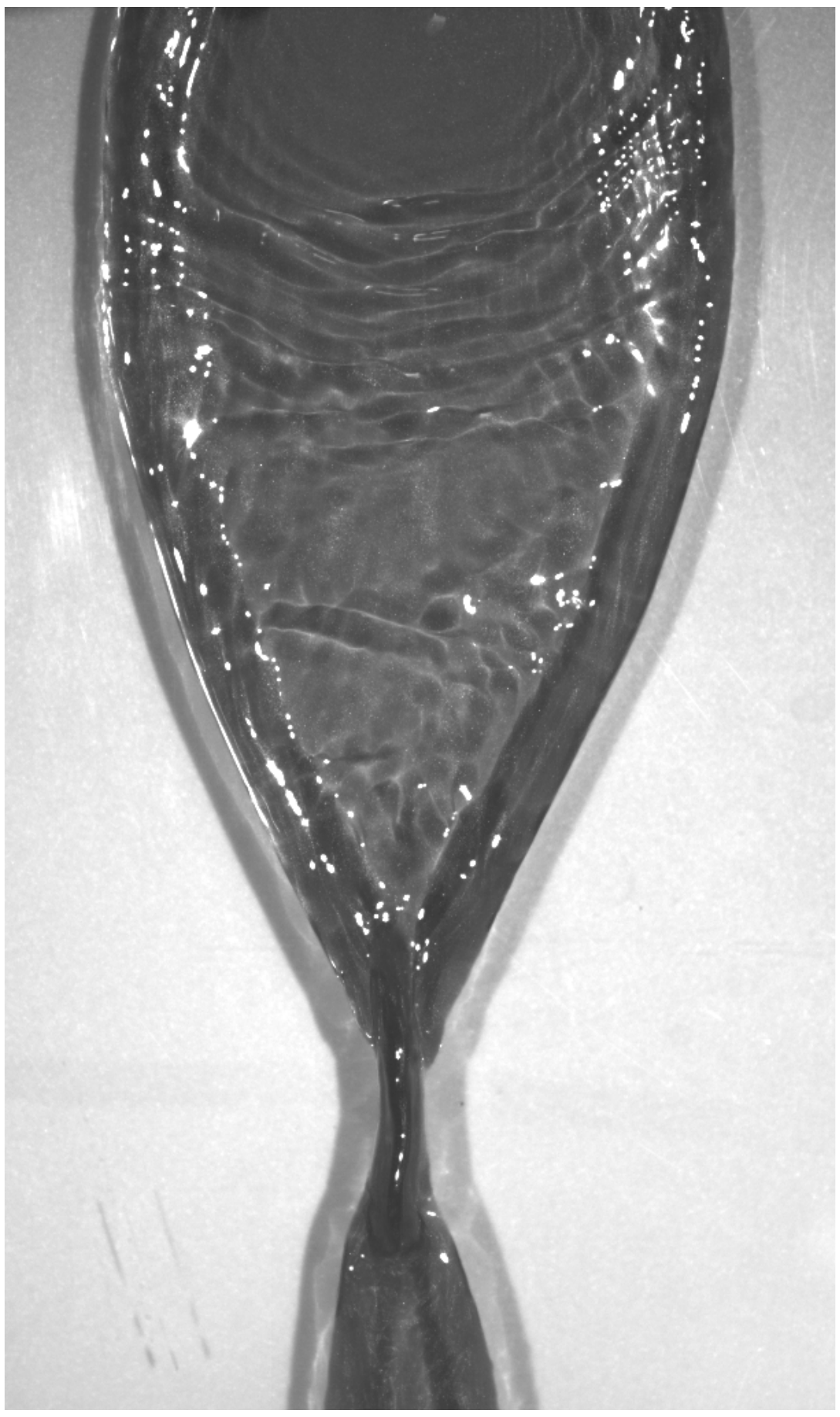

Figure 1 Flow pattern generated by a horizontal liquid jet impinging normally on a vertical wall: (*a*) Schematic, and photographs from PIV testing of (*b*) region around impingement point

(shadow of jet and nozzle visible), and (*c*) draining film and node. Zones and dimensions are described in the text. Experimental conditions: water on Perspex, 20°C, $Q$ = 0.5 L $min^{-1}$ ($\dot{m}$ = 8.33 g $s^{-1}$), O is the point of jet impingement. Dashed line in (a) represents the film jump.

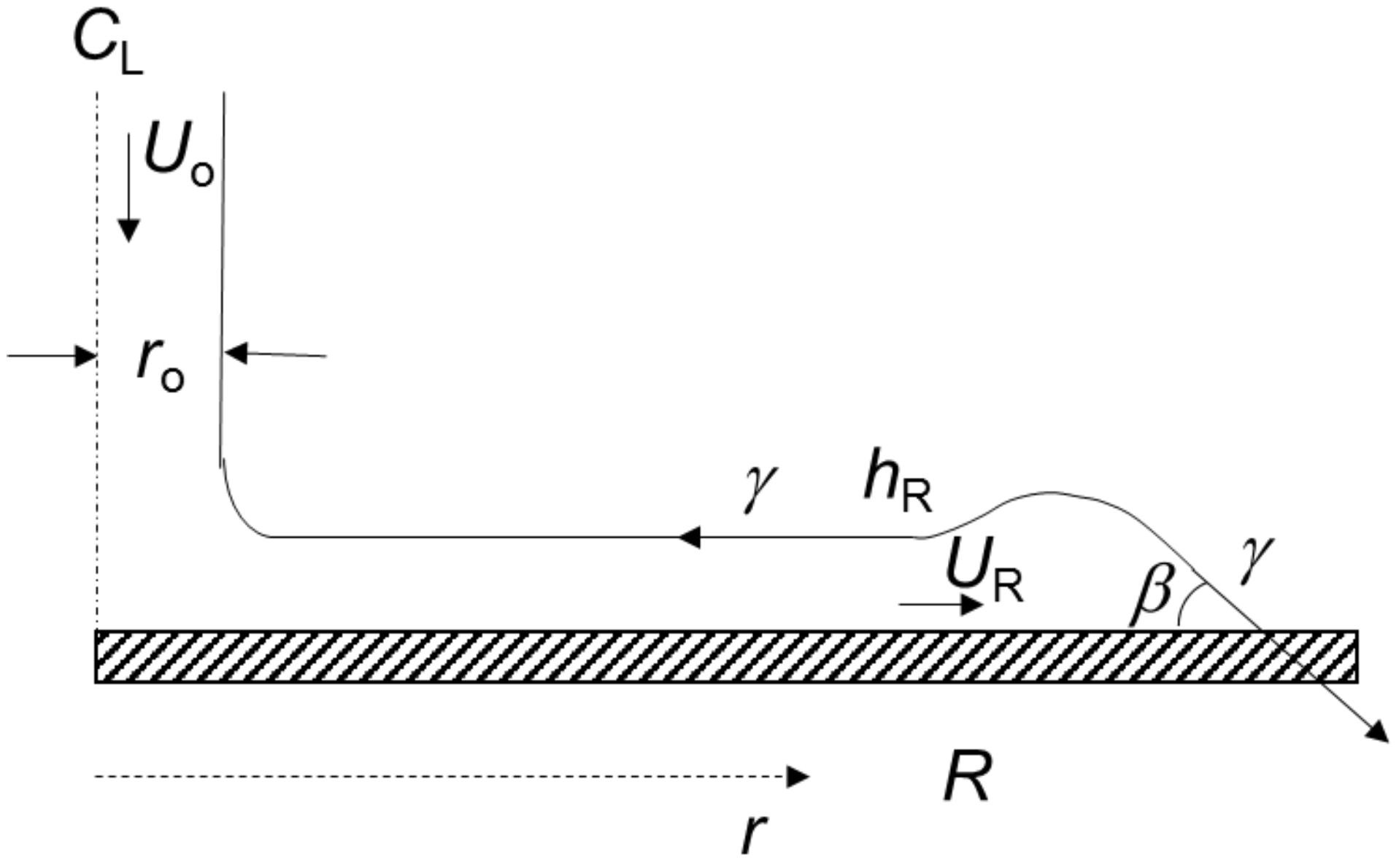

Figure 2 Schematic of a slice through the flow in Zones I and II showing elements in the film jump.

(*a*)

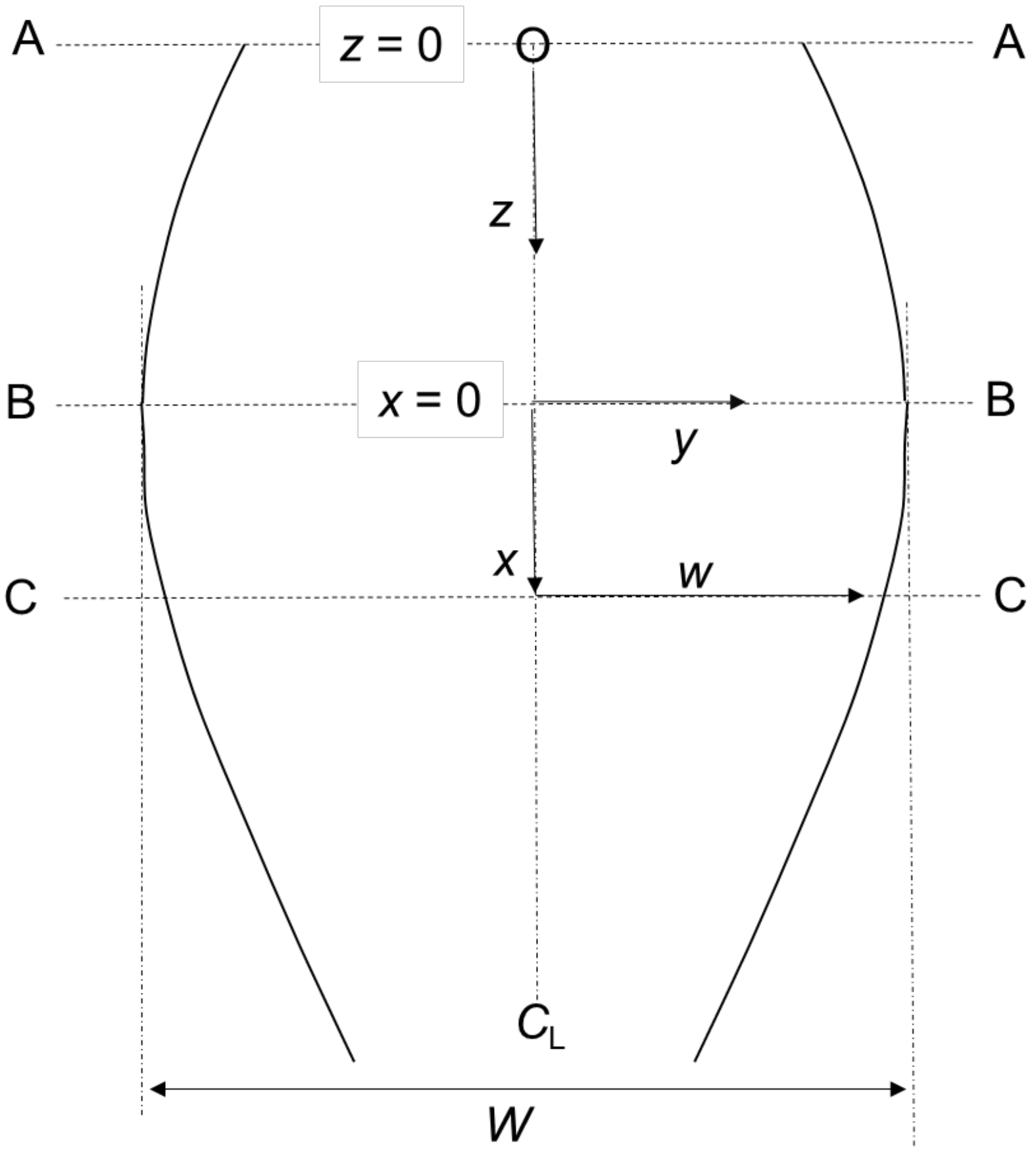

(*b*)

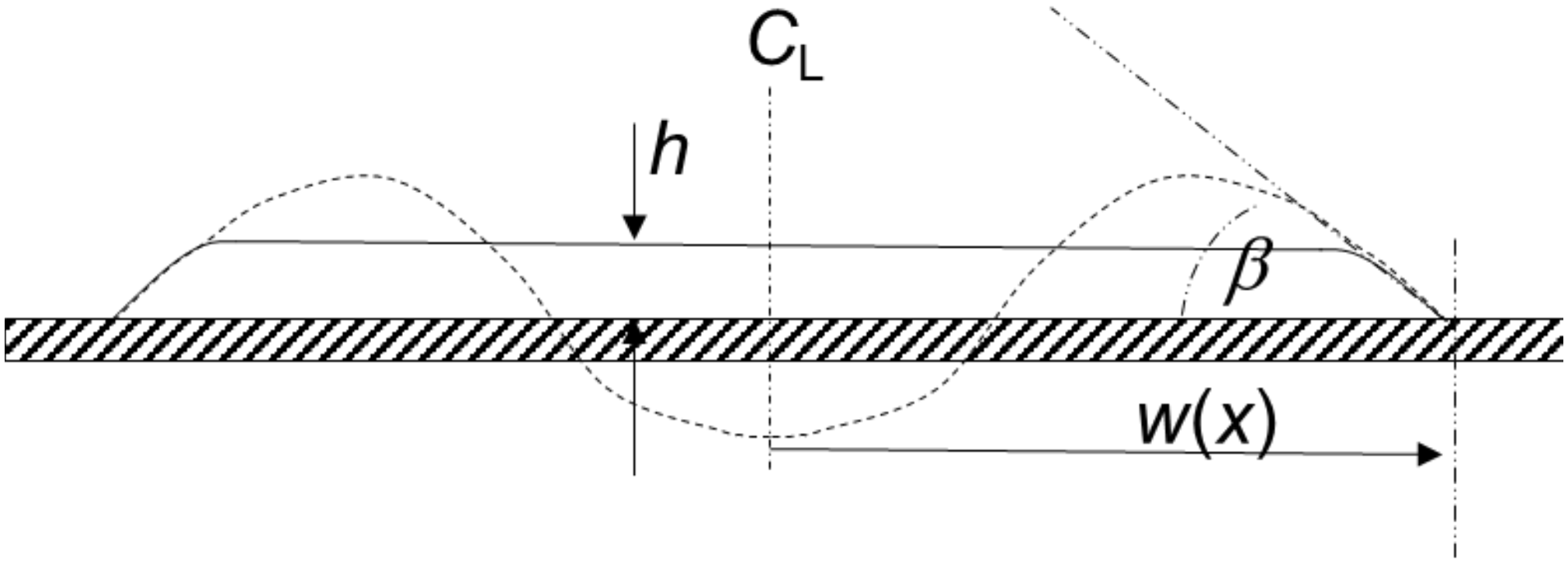

Figure 3. Geometry and dimensions of Zone IV, below BB. (*a*) Coordinates: CC is an arbitrary plane distance *x* downstream of the widest point (BB, Figure 1), where the width is *W*. (*b*) Schematic of film thickness profiles. The dashed line is a sketch of the Mertens *et al*. quartic function (Equation [9]). The solid line represents the simple thin film flow advocated in this work, where the flow is assumed to fall as a Nusselt film. Not to scale.

(*a*)

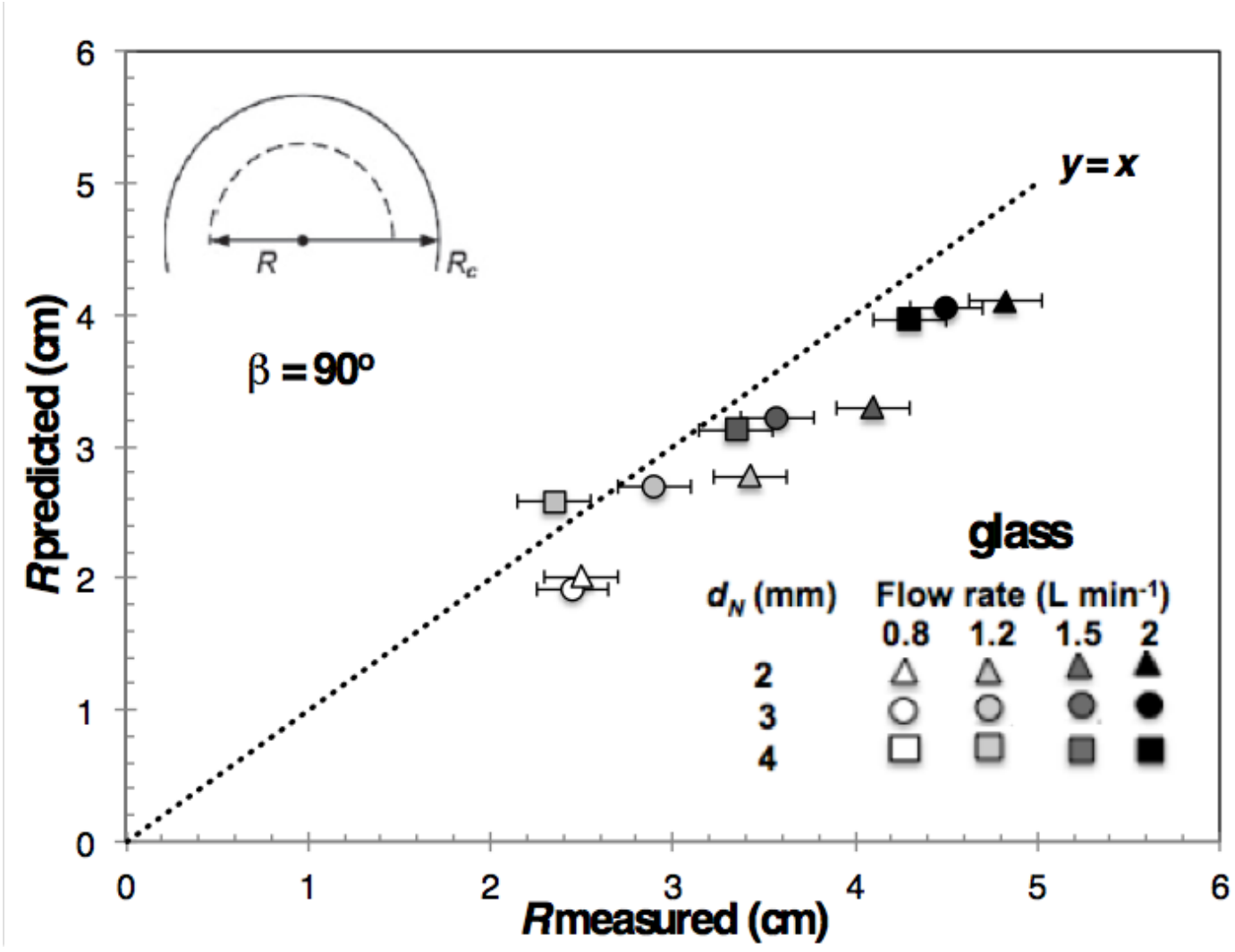

(*b*)

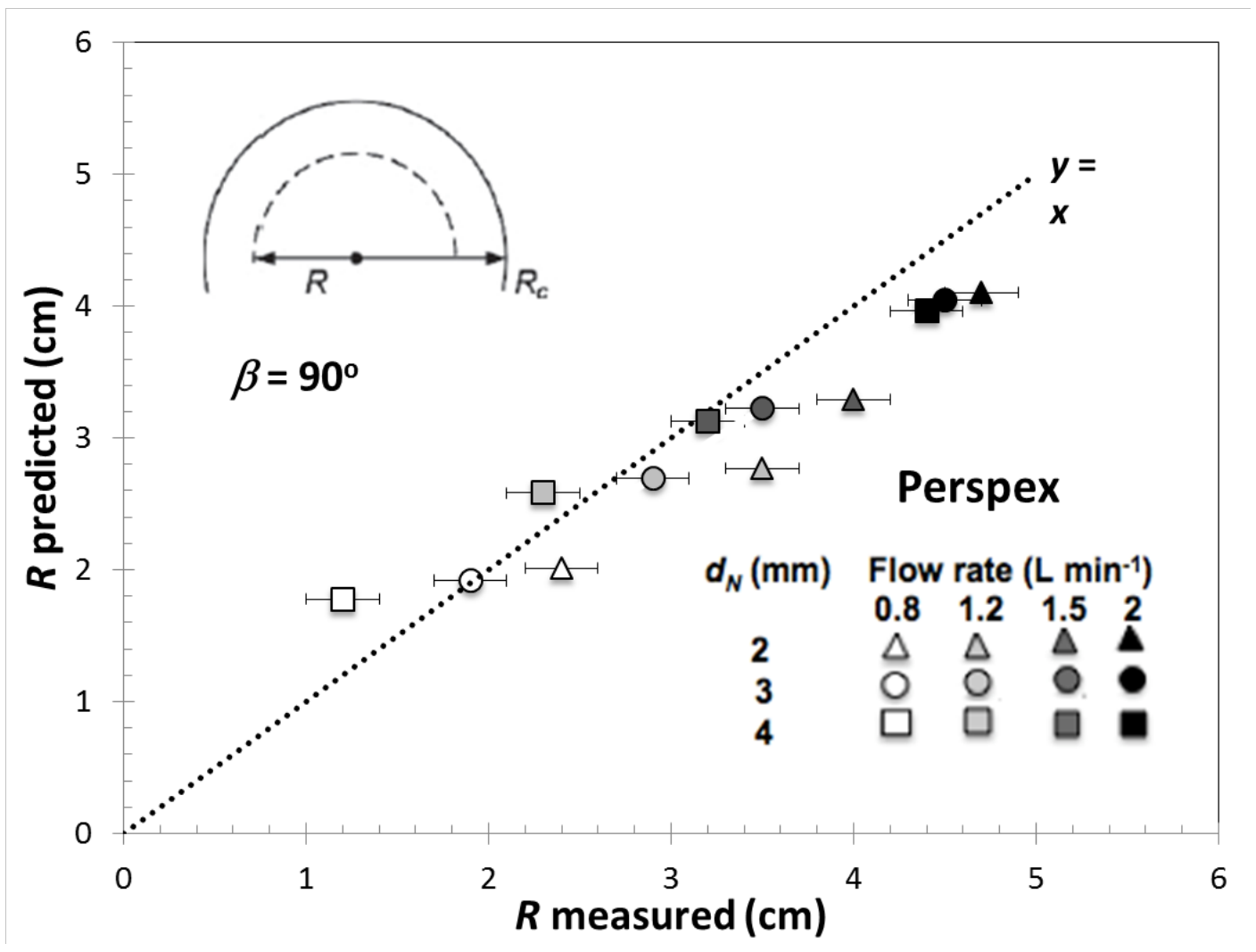

(*c*)

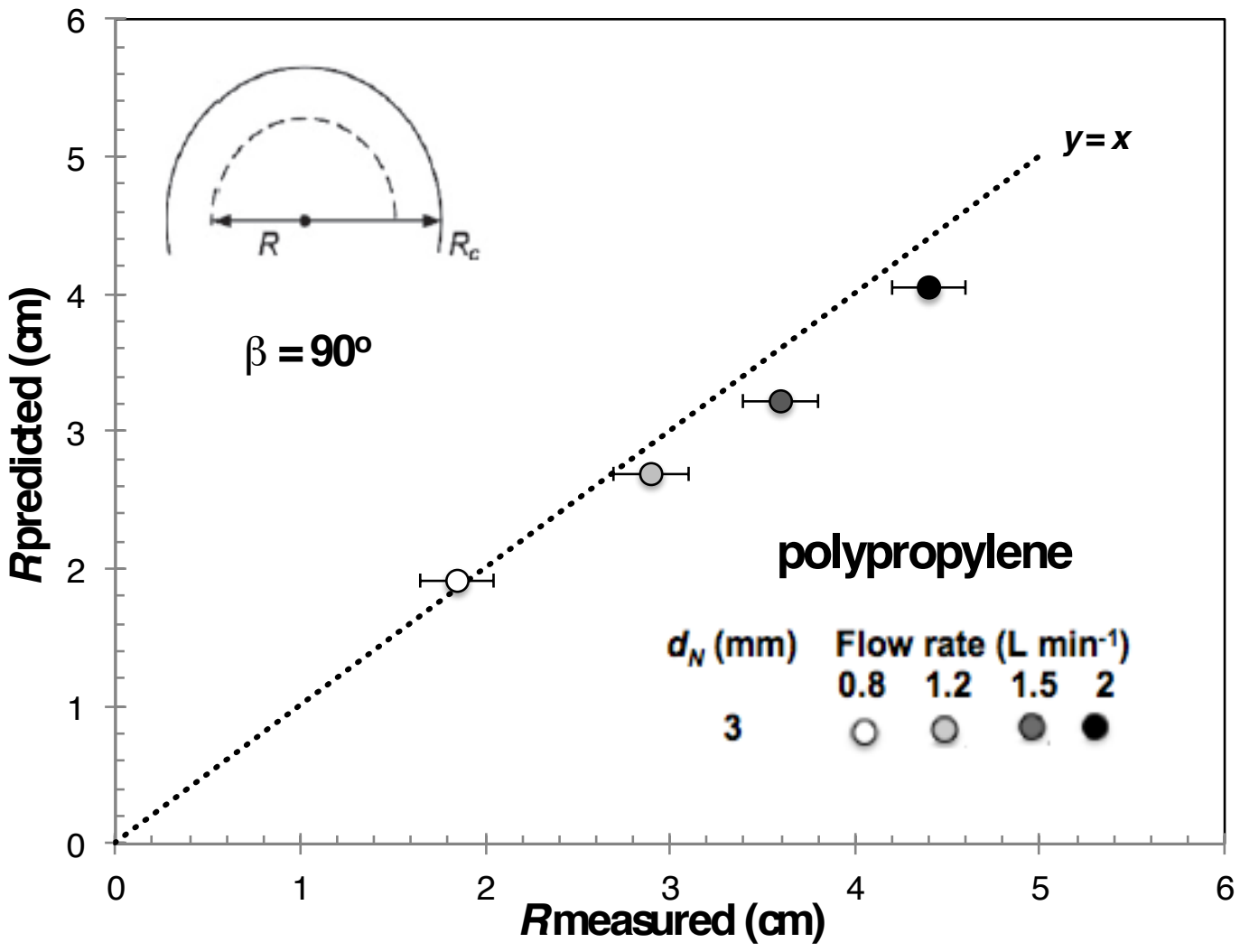

Figure 4 Comparison of predicted film jump radius from Equation [4] with β = 90° for (*a*) glass, (*b*) Perspex, and (*c*) polypropylene surfaces. Legend indicates flow rate and nozzle size. $d_N$ = 4 mm indicates no nozzle (open pipe).

(*a*)

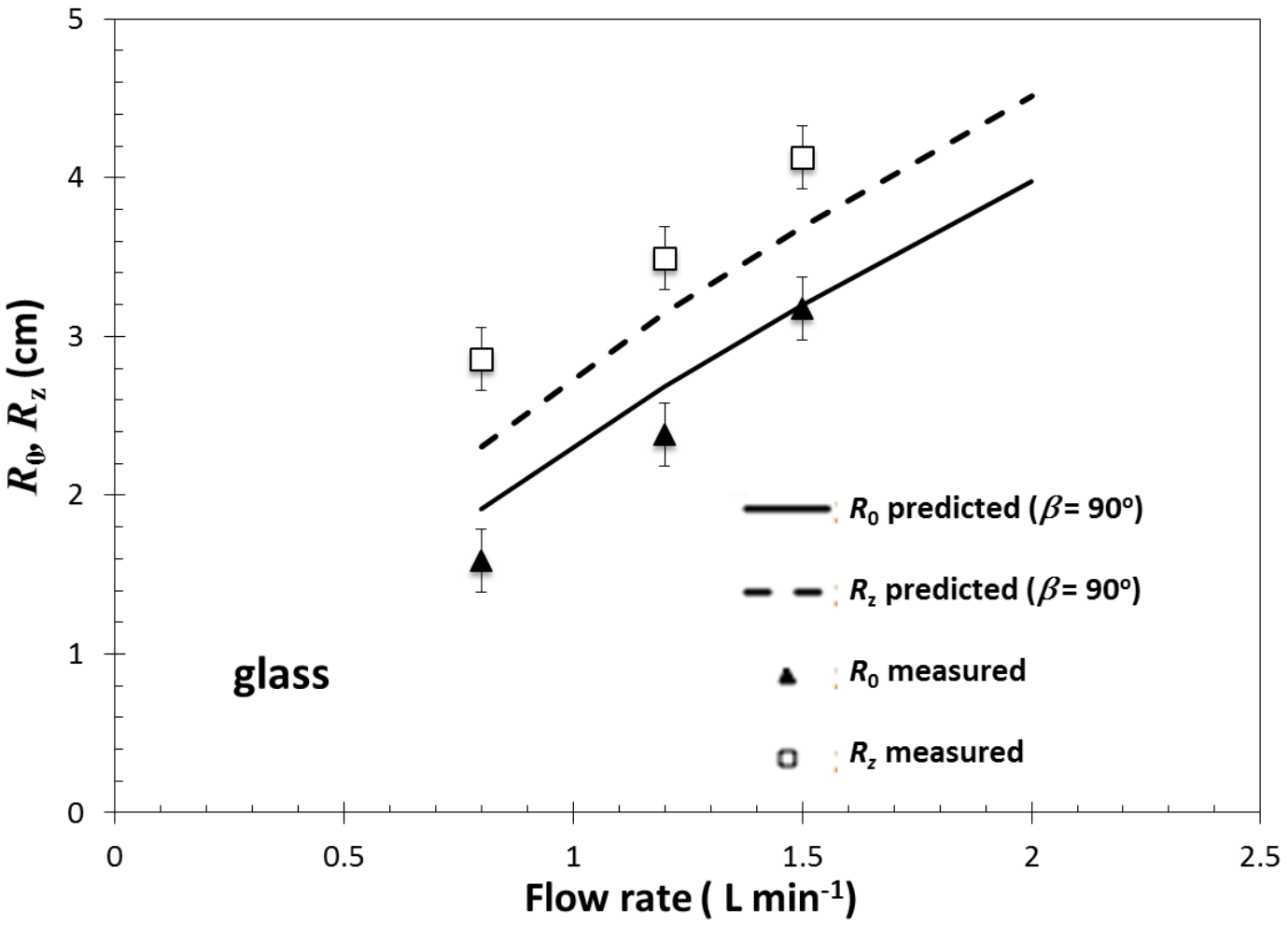

(*b*)

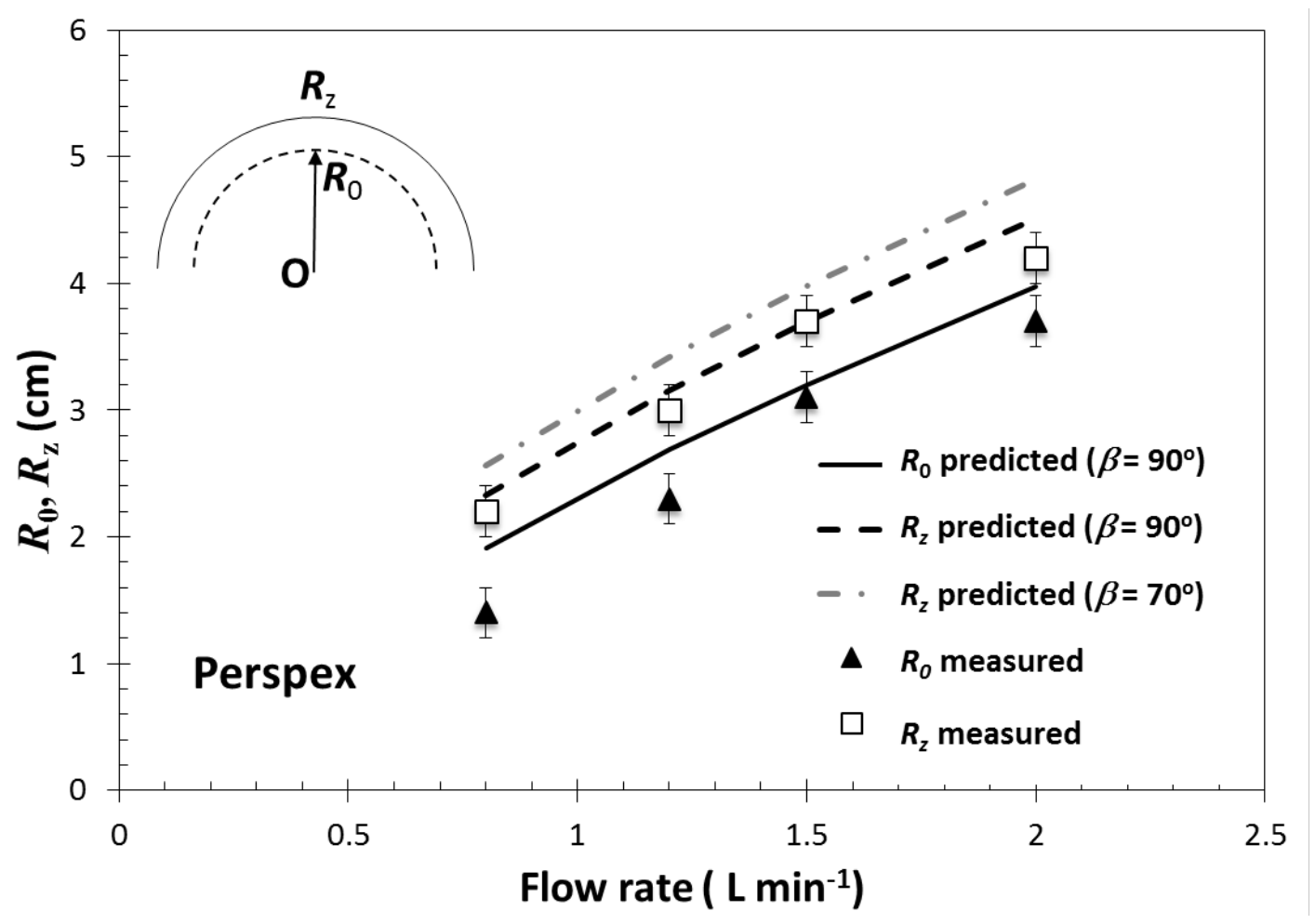

(*c*)

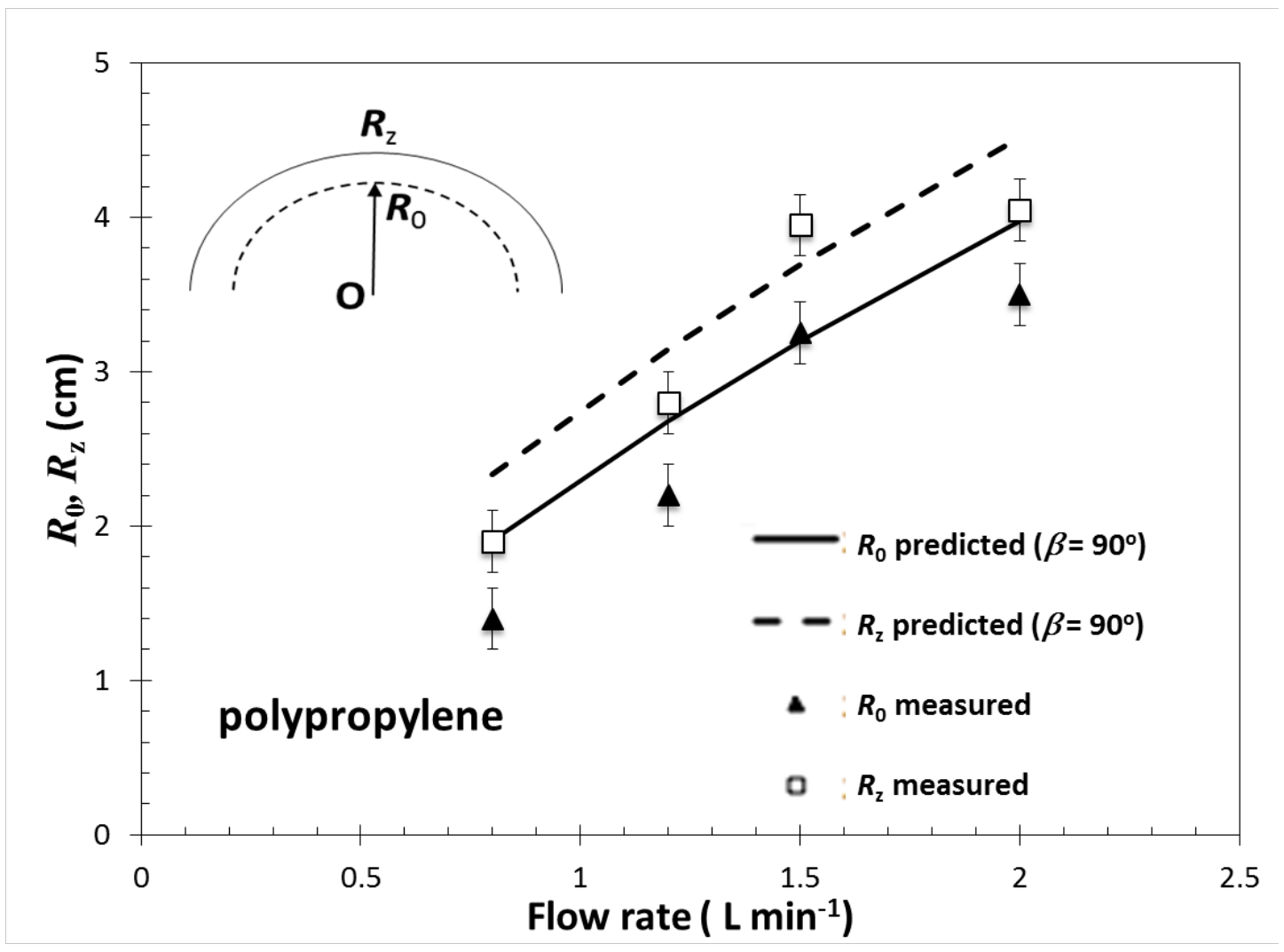

Figure 5 Comparison of measured and predicted extent of the radial flow zone directly above the point of impingement for (*a*) glass, (*b*) Perspex, and (*c*) polypropylene using an effective contact angle of 90°. for Perspex with $\beta = 70°$ also shown. The nozzle diameter, $d_N$, is 3 mm. The predicted values of $R_0$ were obtained from Equations [2] and [3]; $R_z$ was estimated using Equation [8].

(a)

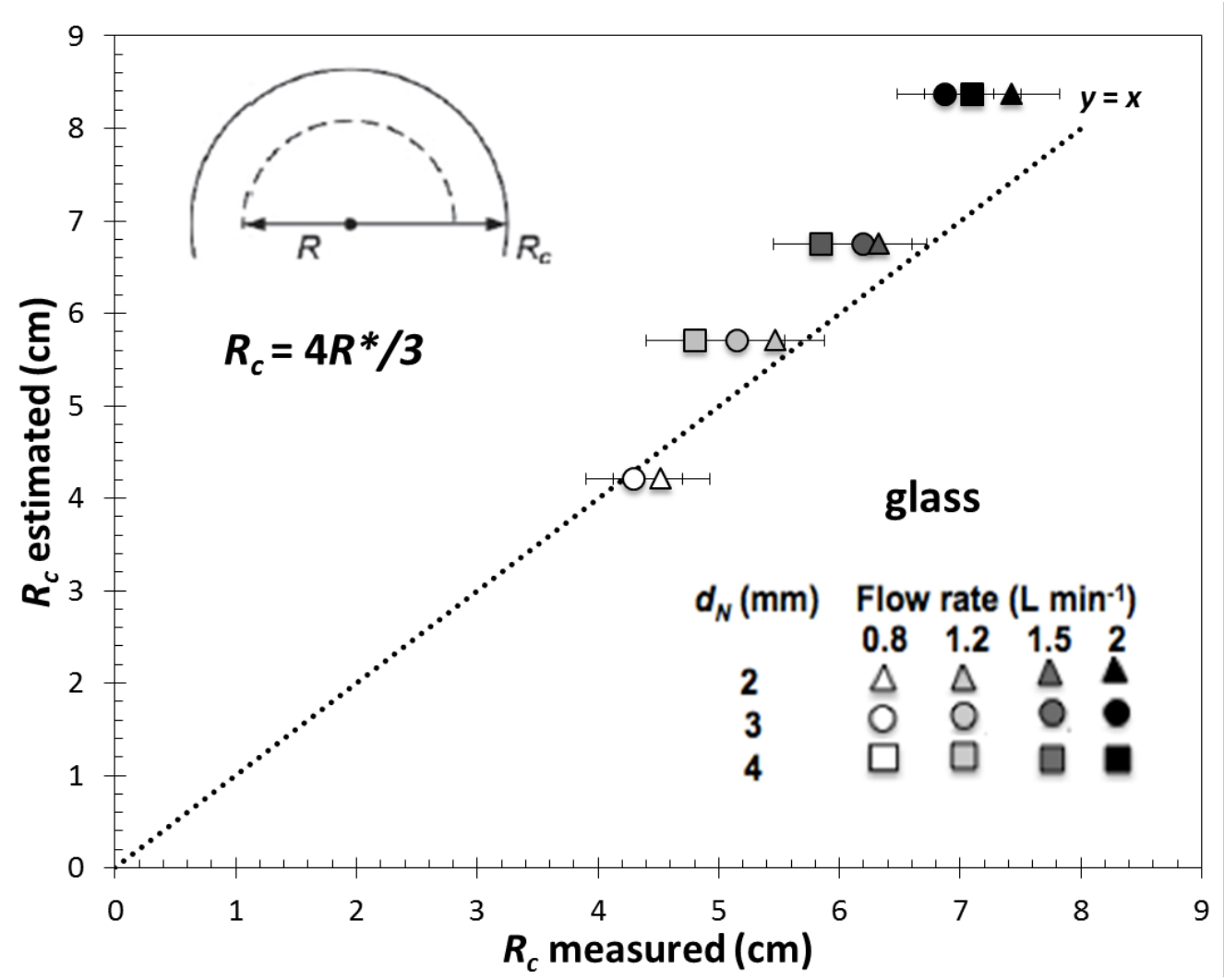

(b)

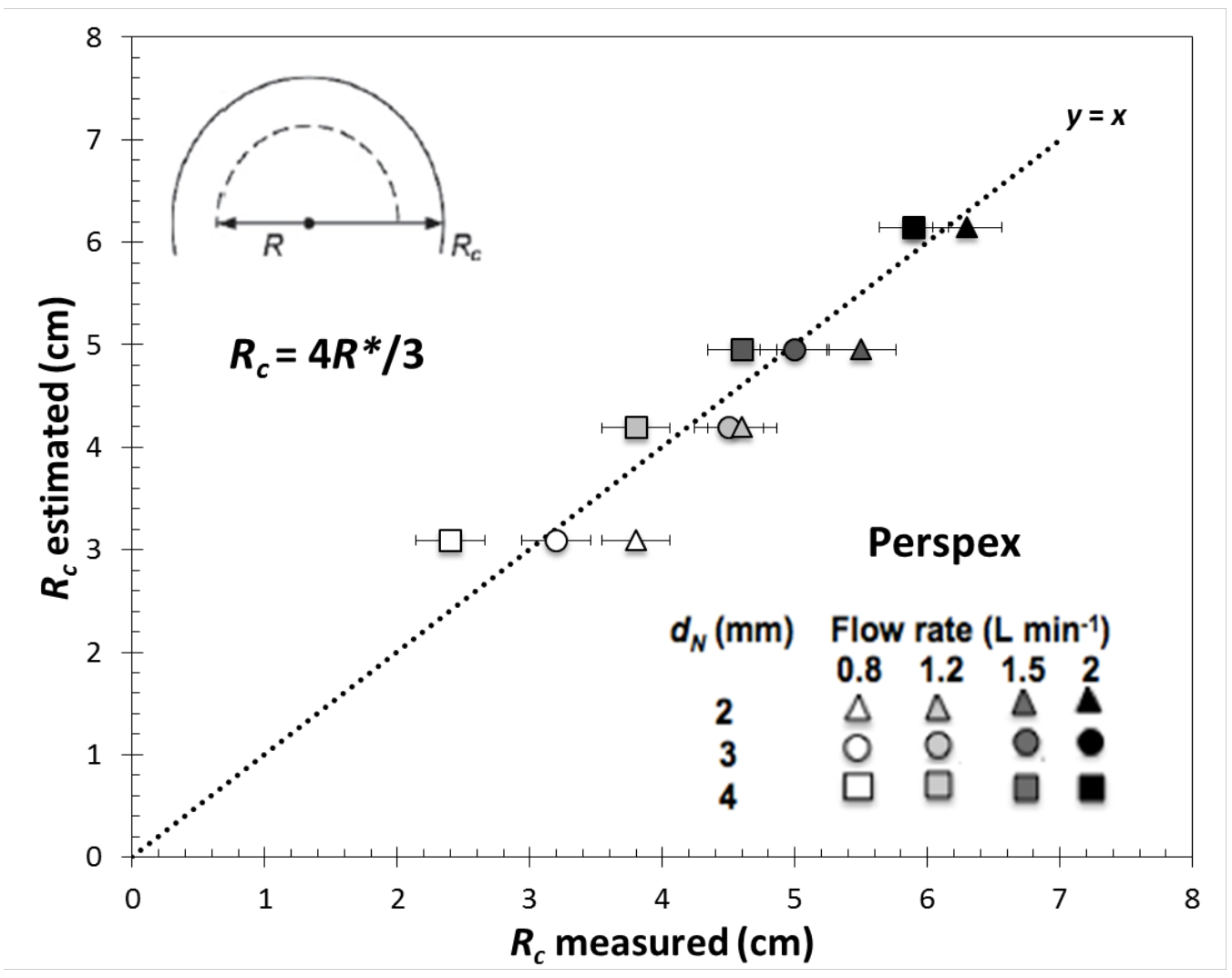

(c)

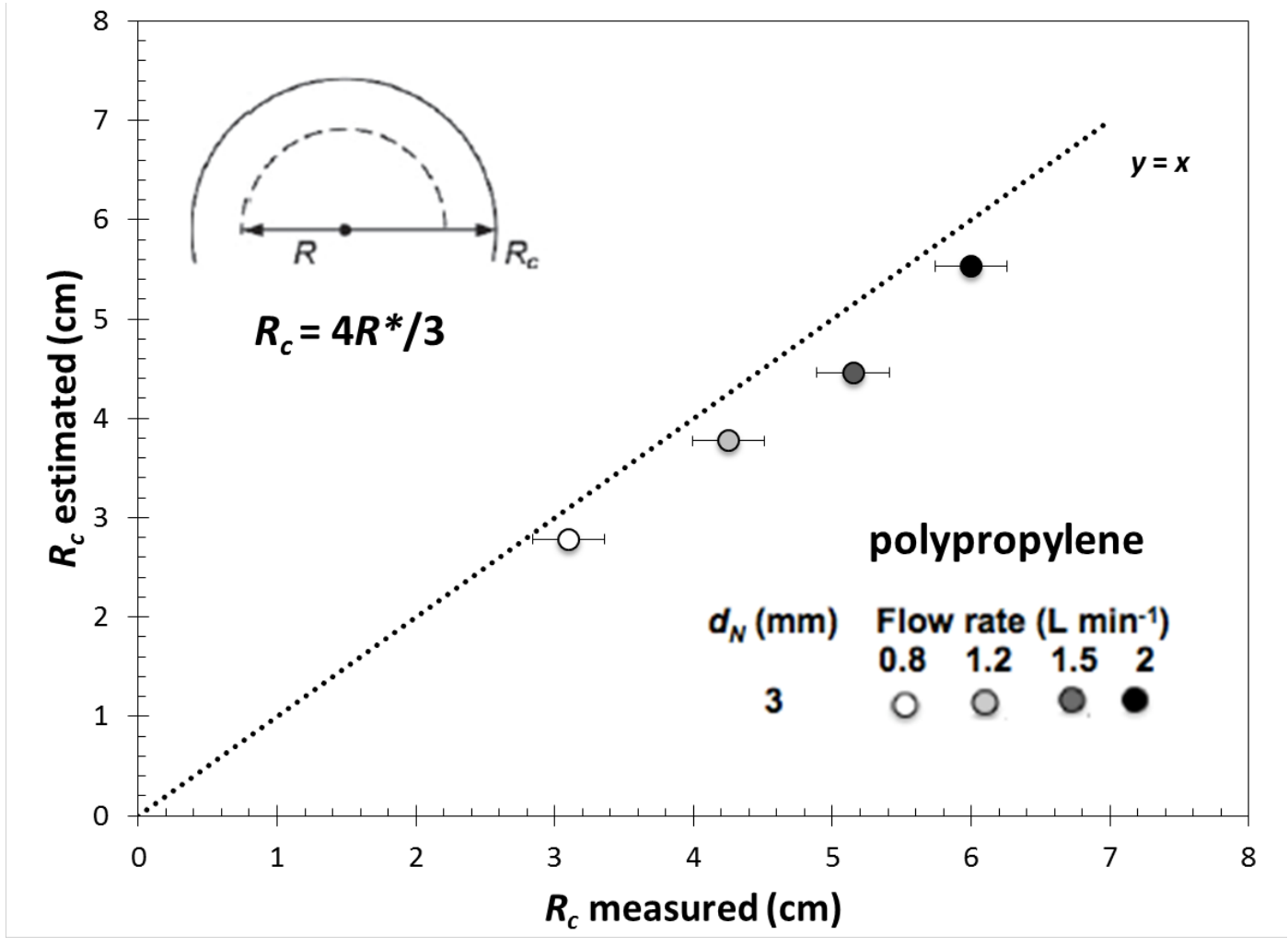

Figure 6 Measured half-width of wetted region on AA, $R_c$, plotted against $R_c = 4/3R^*$, where $R^*$ is predicted using Equation [4] for (*a*) glass; (*b*) Perspex; and (*c*) polypropylene. $d_N$ = 4 mm indicates no nozzle (open pipe).

(*a*) (*b*)

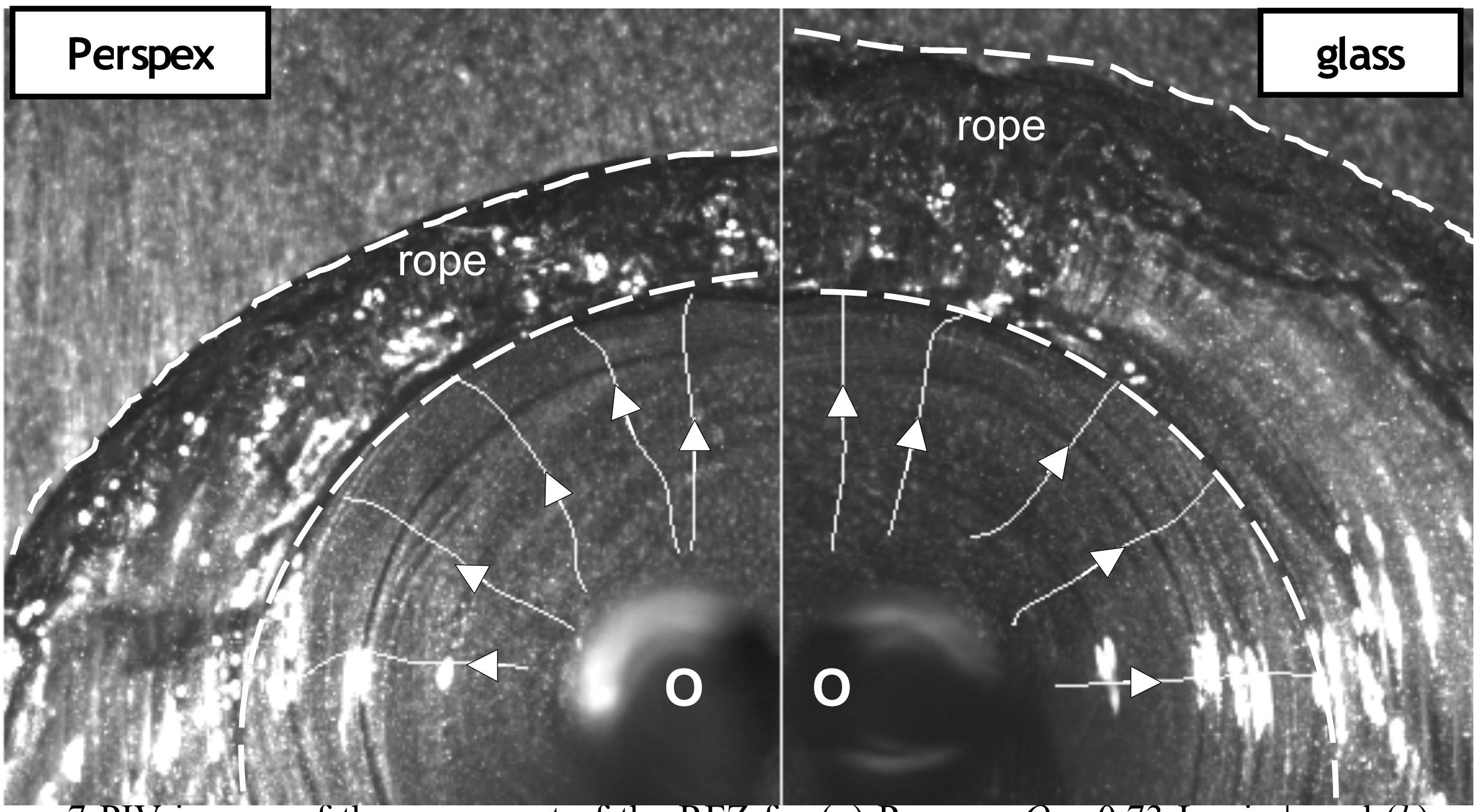

Figure 7 PIV images of the upper part of the RFZ for (*a*) Perspex, $Q$ = 0.73 L min$^{-1}$, and (*b*) glass, $Q$ = 0.71 L min$^{-1}$. $d_N$ = 2 mm. Nozzle obscures point of impingement O. Dashed lines show boundary of rope (Zone II). White loci with arrows show instantaneous streamlines identified by PIV.

(*a*)

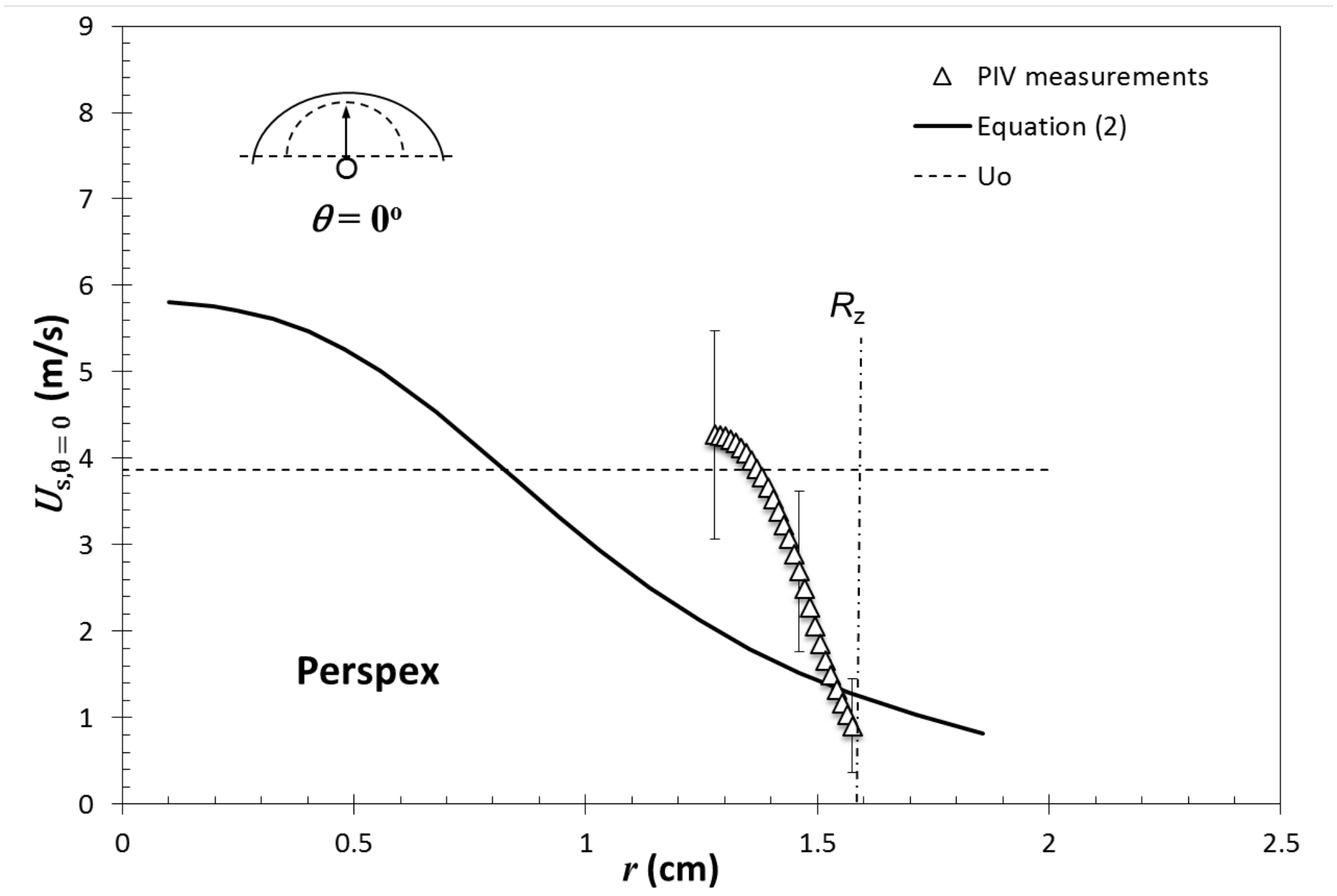

(*b*)

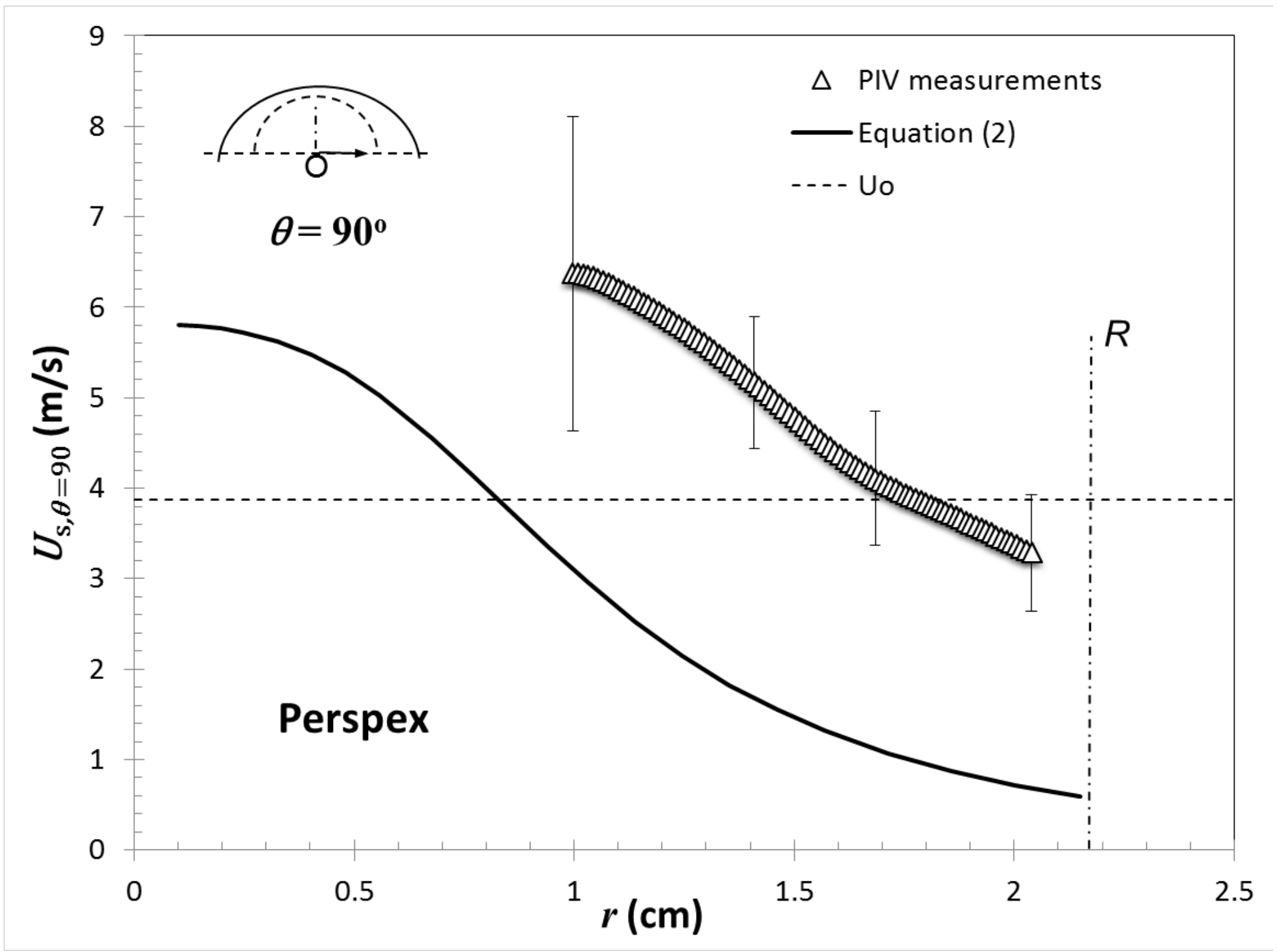

Figure 8 Comparison between the predicted surface velocity and measured values for the flow pattern in Figure 7(*a*), $Q$ = 0.73 L min$^{-1}$ on Perspex, with a 2 mm nozzle: (*a*) vertical direction; (*b*) horizontal direction. Error bars on PIV measurements show experimental uncertainty for selected positions. Dashed horizontal line shows jet velocity, $U_o$.

(*a*)

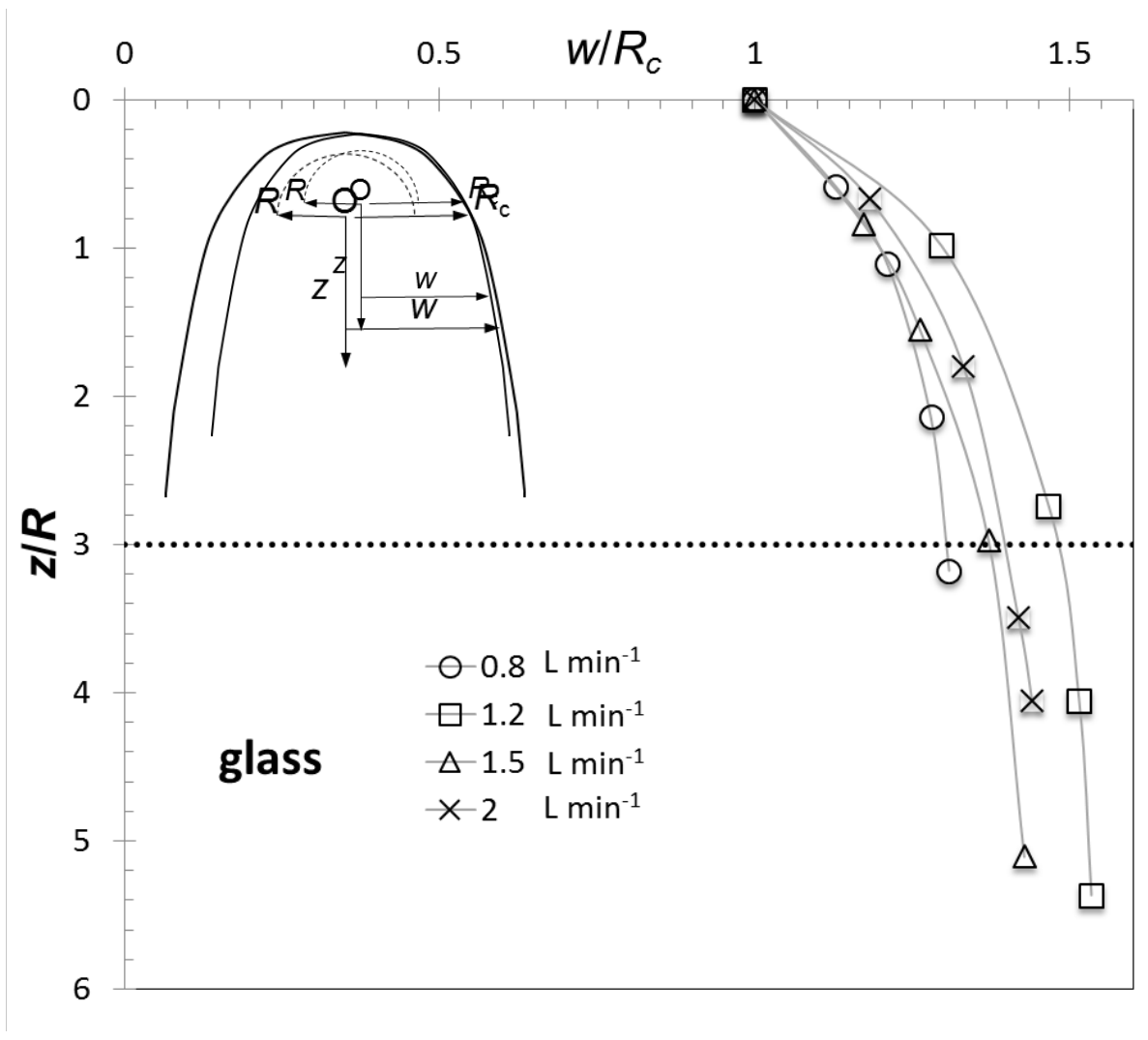

(b)

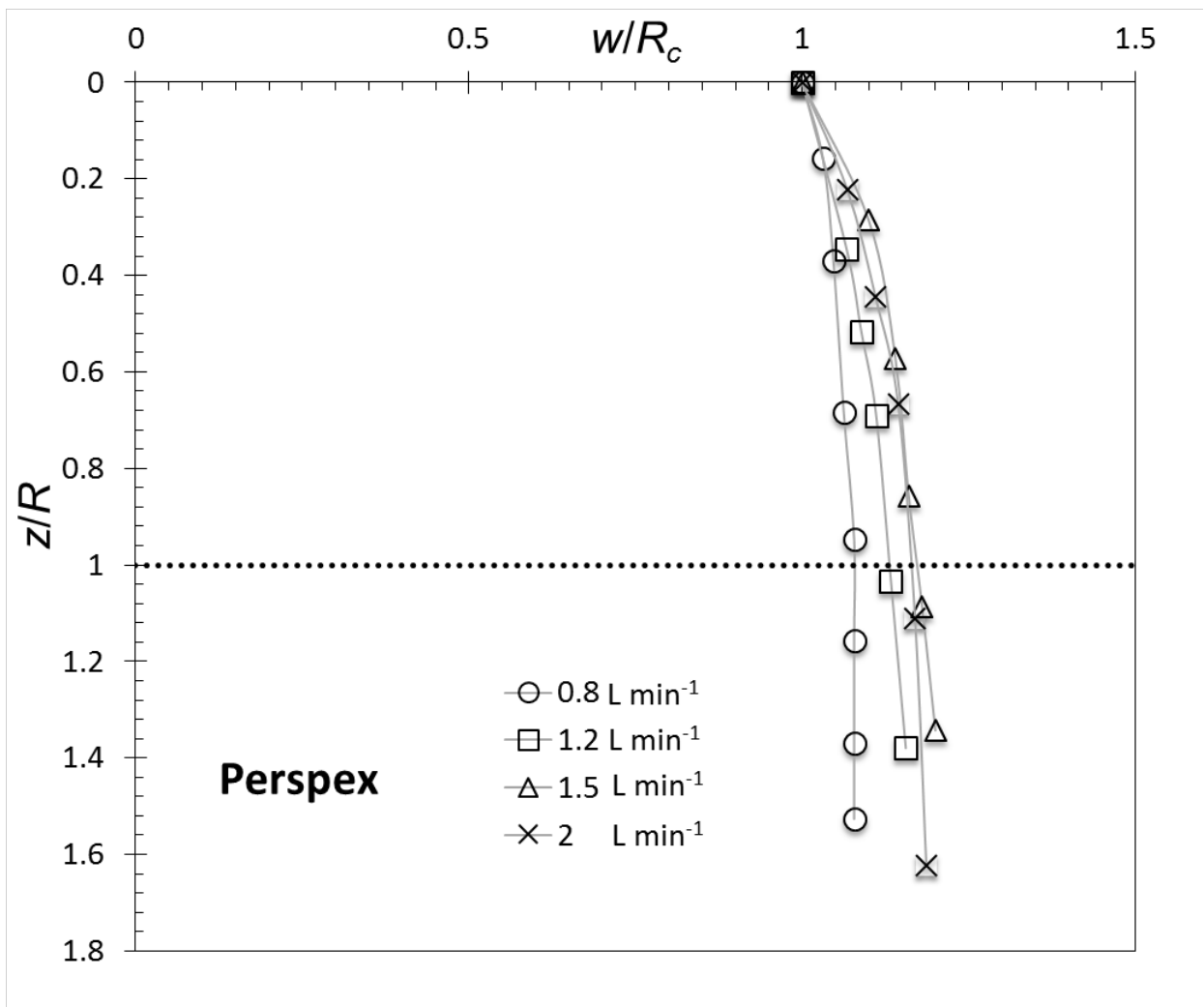

(*c*)

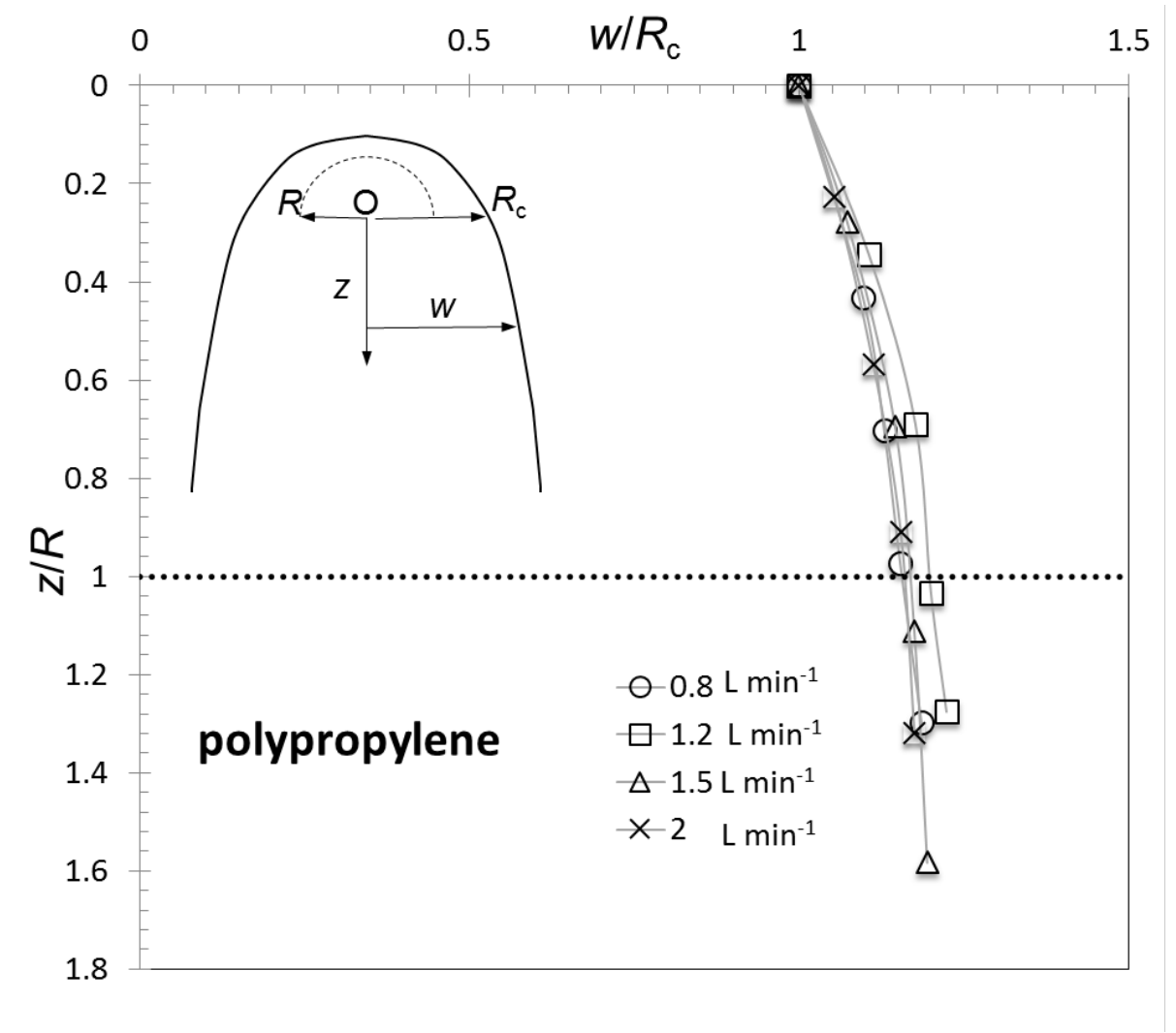

Figure 9 Shape of draining film in zone III for water on (*a*) glass; (*b*) Perspex; (*c*) polypropylene. 3 mm nozzle, flow rate shown in legend. Co-ordinates presented in dimensionless form, $z/R$ and $w/R_c$, where $R$ and $R_c$ are experimental measurements.

(*a*)

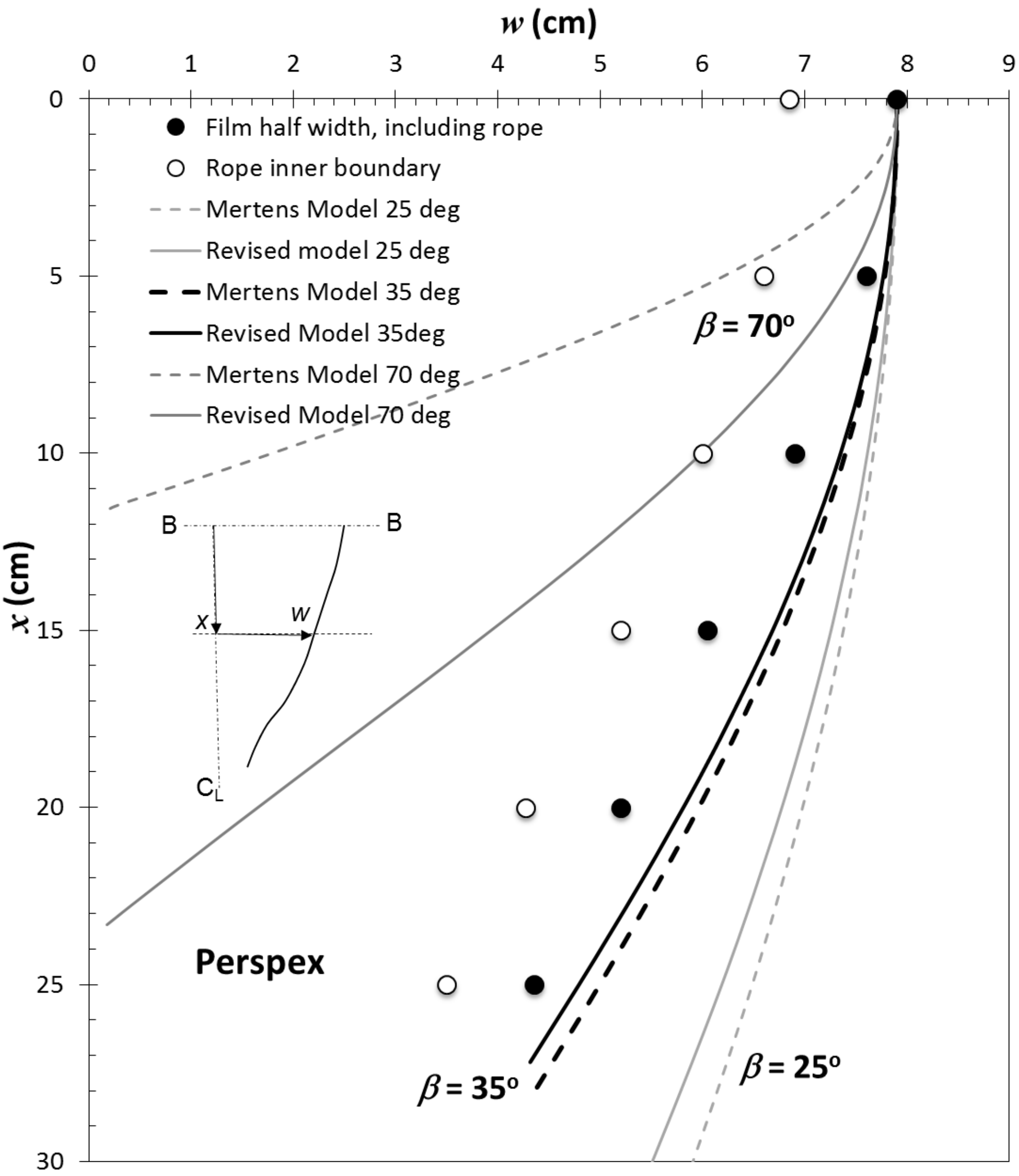

Figure 10 Comparison of wetted region half width, *w*, including rope, in Zone IV on Perspex, $d_N$ = 3 mm. The four diagrams (*a*)-(*d*) show the effect of flow rate: (*a*) $Q$ = 2.3 L min$^{-1}$; (*b*) $Q$ = 1.5 L min$^{-1}$; (*c*) $Q$ = 2.0 L min$^{-1}$, and (*d*) $Q$ = 4.0 L min$^{-1}$. The data points show measurements of film half width (•) and width to inside of rope (o). Loci predicted by the (*i*) Mertens Model and (*ii*) Revised Model, with $\beta_{fit}$ = 35° shown on (*a-d*): (*a*) also shows the predictions for $\beta = \beta_{receding}$ = 25° and $\beta = \beta_{advancing}$ = 70°.

(*b*)

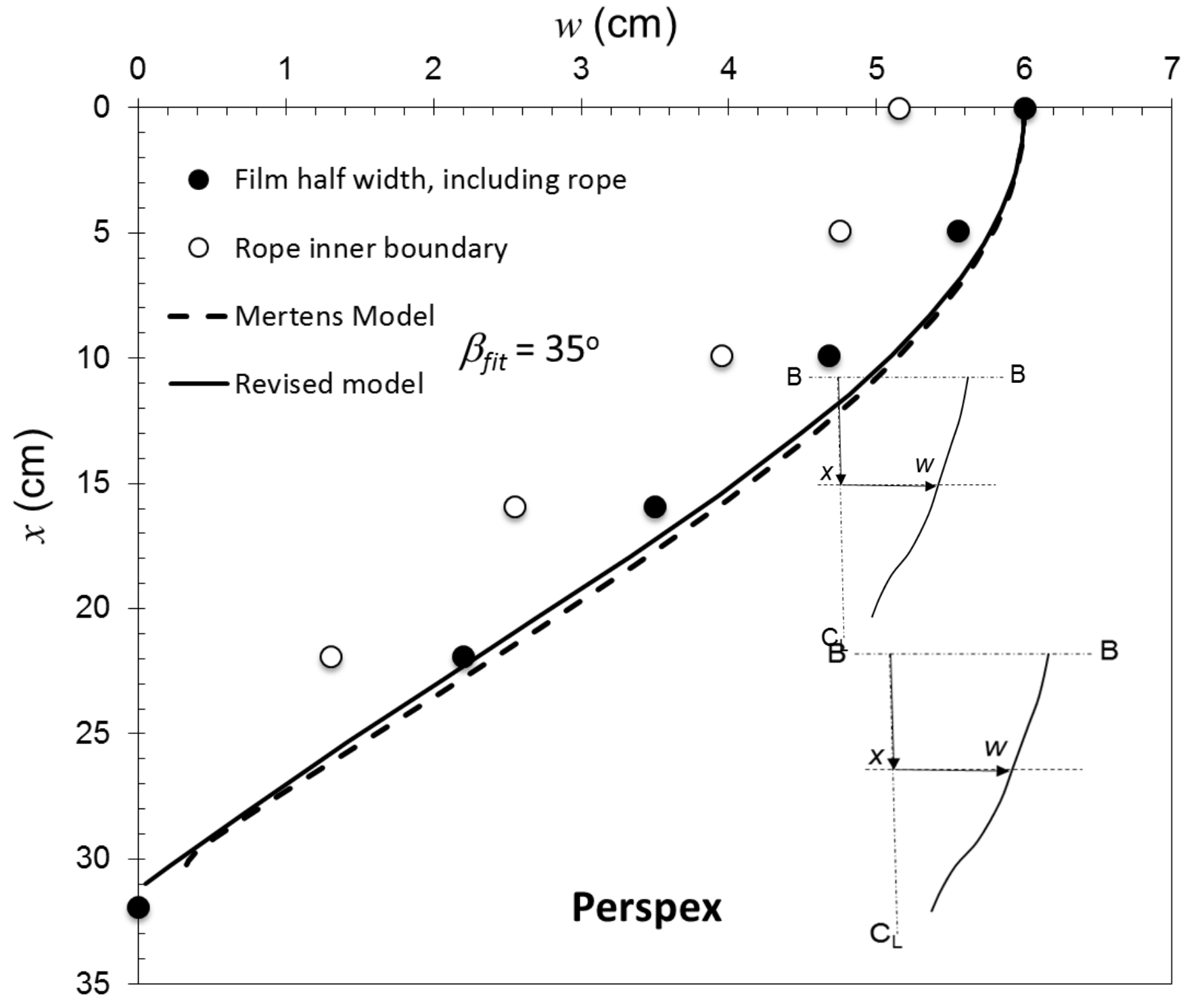

(*c*)

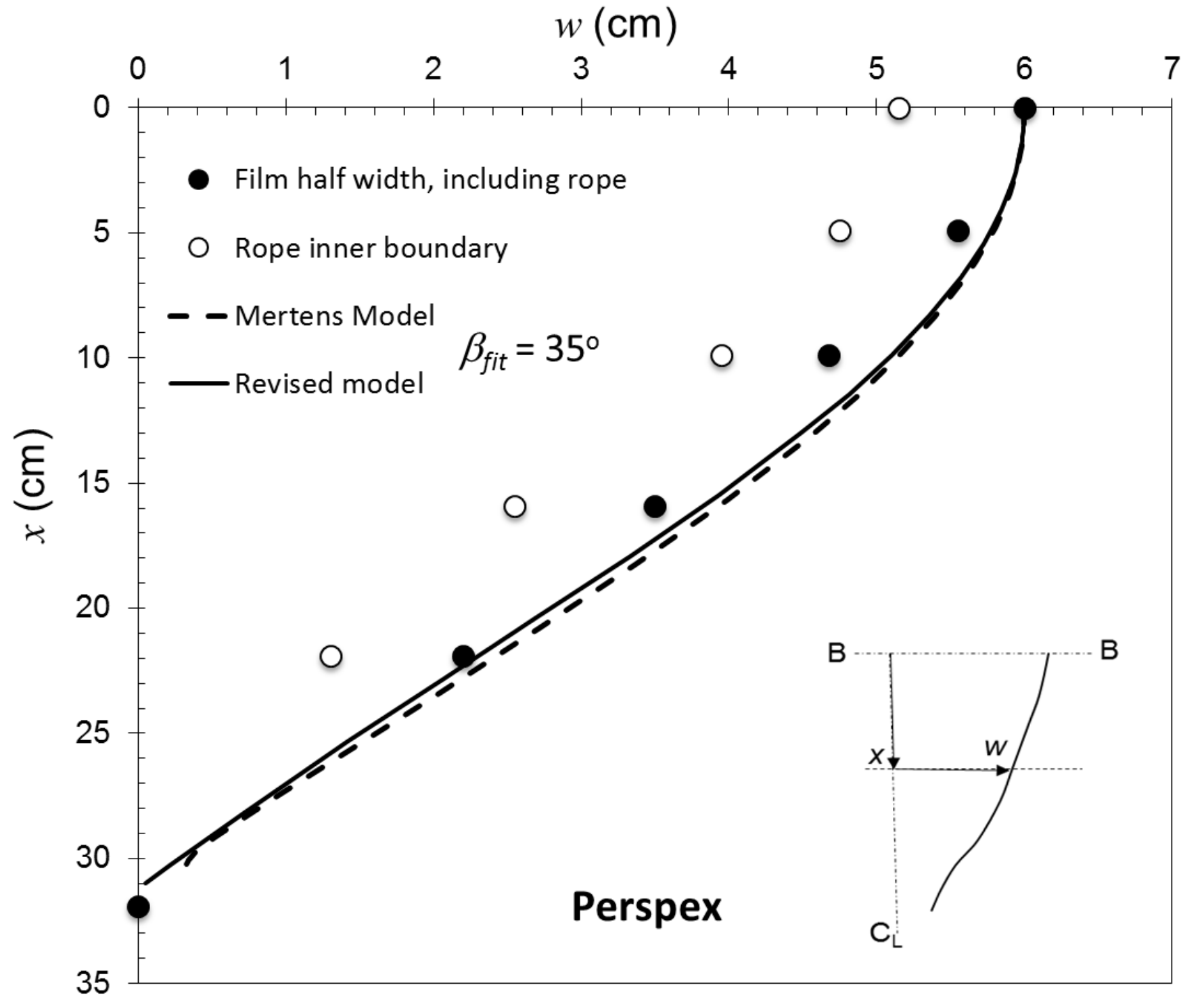

Figure 10 Comparison of wetted region half width, *w*, including rope, in Zone IV on Perspex, $d_N$ = 3 mm. The four diagrams (*a*)-(*d*) show the effect of flow rate: (*a*) $Q$ = 2.3 L min$^{-1}$; (*b*) $Q$ = 1.5 L min$^{-1}$; (*c*) $Q$ = 2.0 L min$^{-1}$, and (*d*) $Q$ = 4.0 L min$^{-1}$. The data points show measurements of film half width (•) and width to inside of rope (o). Loci predicted by the (*i*) Mertens Model and (*ii*) Revised Model, with $\beta_{fit}$ = 35° shown on (*a*-*d*): (*a*) also shows the predictions for $\beta = \beta_{receding}$ = 25° and $\beta = \beta_{advancing}$ = 70°.

(*d*)

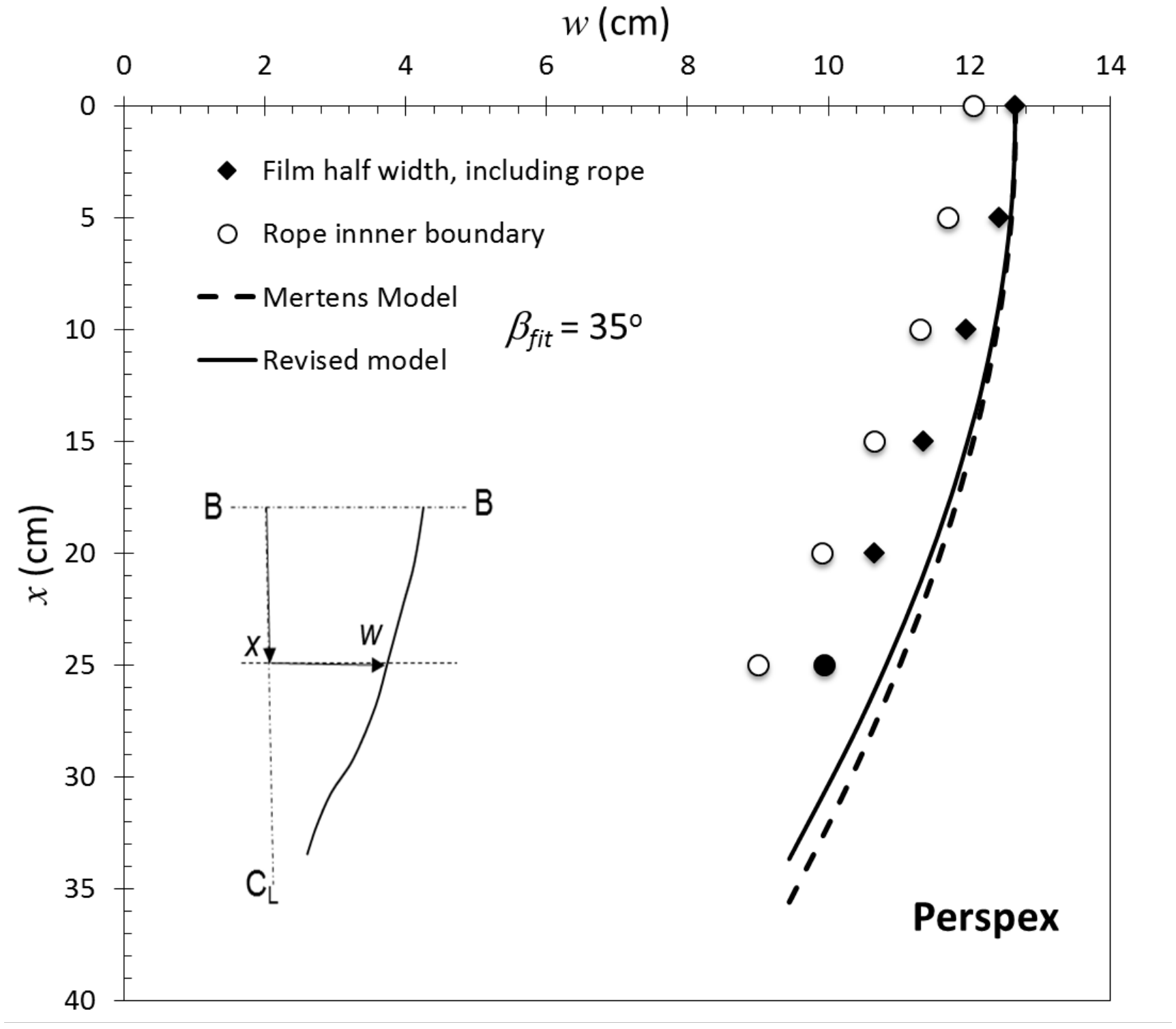

Figure 10 Comparison of wetted region half width, *w*, including rope, in Zone IV on Perspex, $d_N$ = 3 mm. The four diagrams (*a*)-(*d*) show the effect of flow rate: (*a*) $Q$ = 2.3 L min$^{-1}$; (*b*) $Q$ = 1.5 L min$^{-1}$; (*c*) $Q$ = 2.0 L min$^{-1}$, and (*d*) $Q$ = 4.0 L min$^{-1}$. The data points show measurements of film half width (•) and width to inside of rope (o). Loci predicted by the (*i*) Mertens Model and (*ii*) Revised Model, with $\beta_{fit}$ = 35° shown on (*a*-*d*): (*a*) also shows the predictions for $\beta = \beta_{receding}$ = 25° and $\beta = \beta_{advancing}$ = 70°.

(*a*, *i*) (*a*, *ii*)

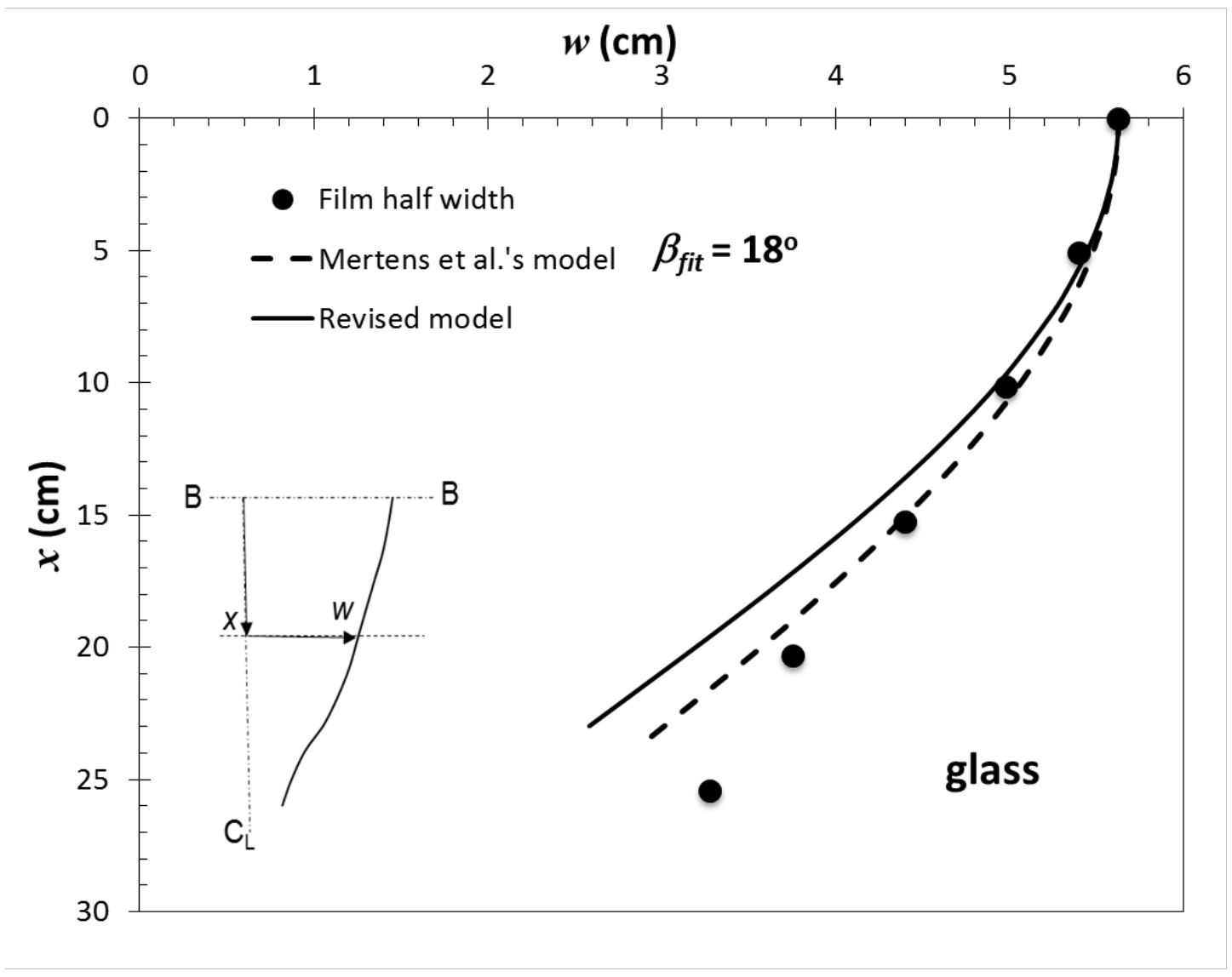

w (cm)
0 1 2 3 4 5 6
0 5 10 15 20 25 30
x (cm)
Film half width
Mertens et al.'s model
Revised model
βfit = 18o
B
B
x
w
CL
glass

(*b*, *i*) (*b*, *ii*)

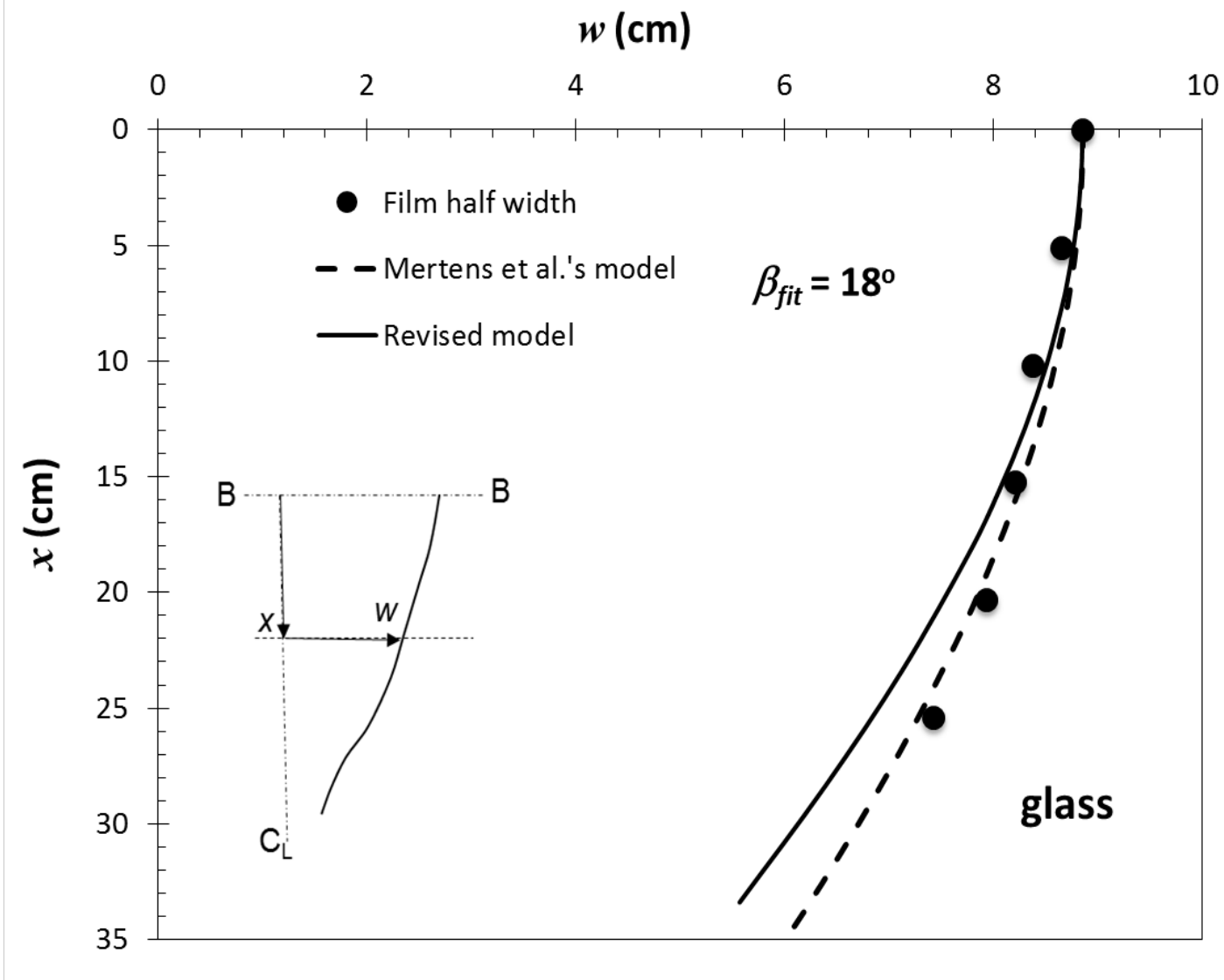

w (cm)
0 2 4 6 8 10
0 5 10 15 20 25 30 35
x (cm)
Film half width
Mertens et al.'s model
Revised model
βfit = 18o
B
B
x
w
CL
glass

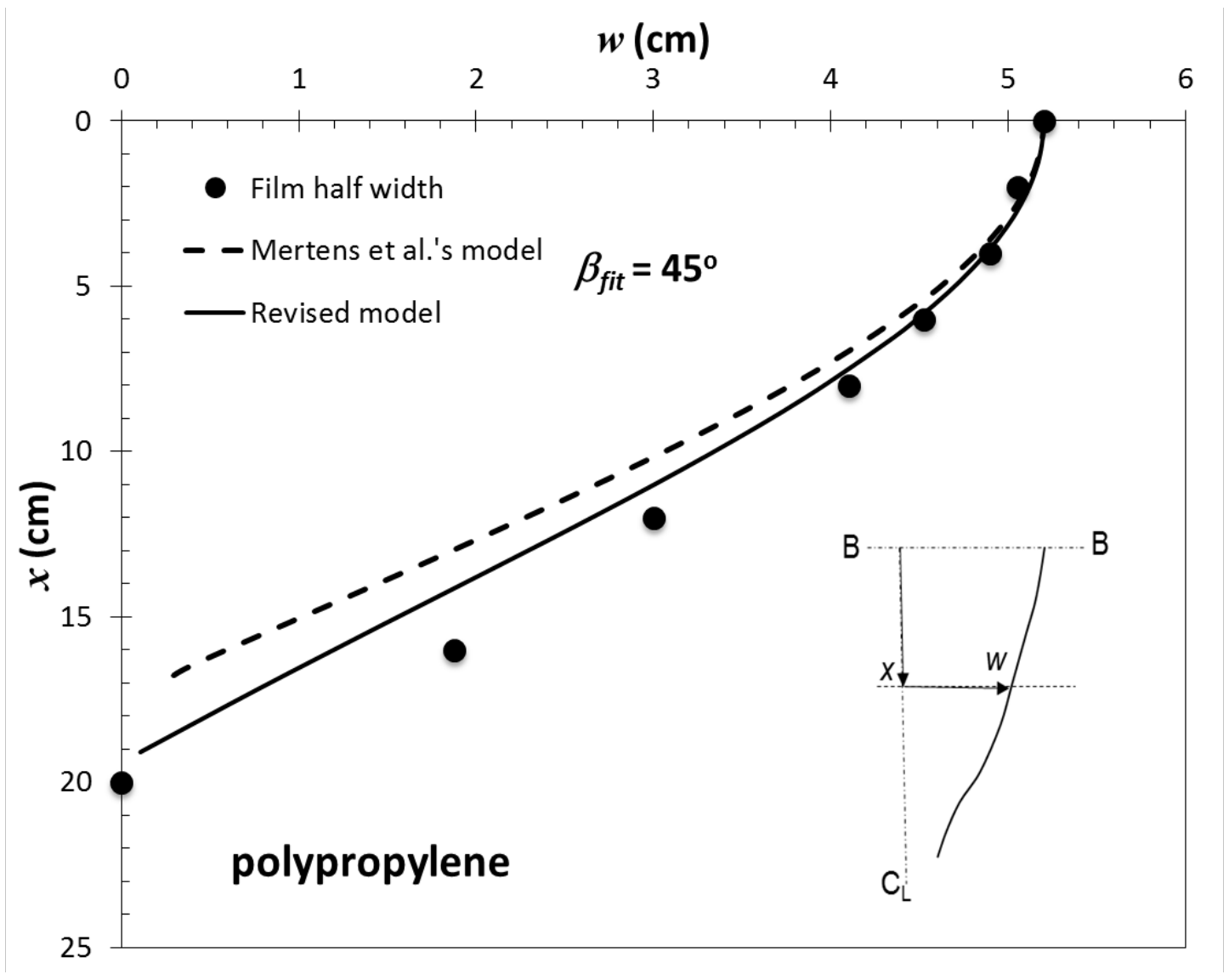

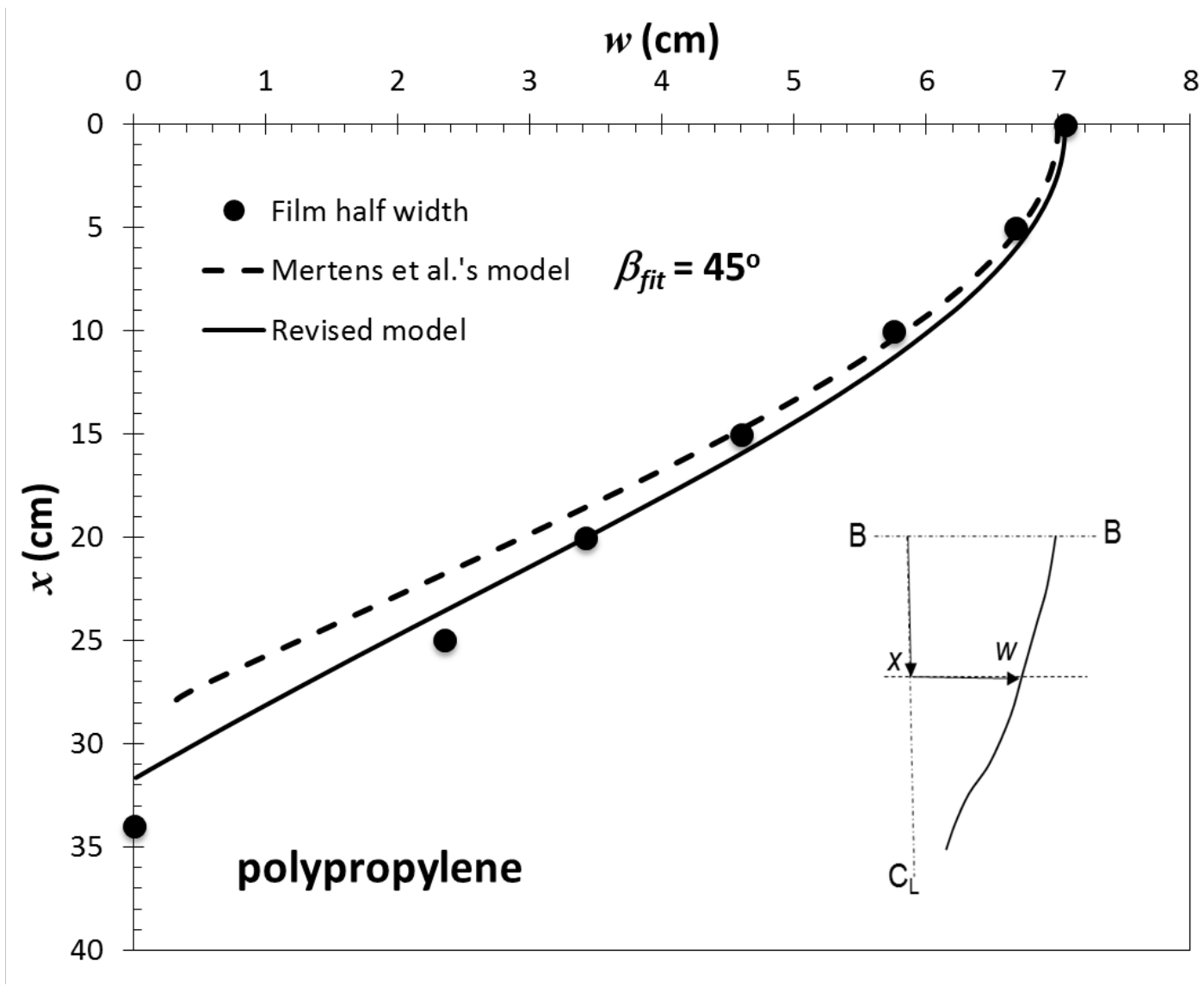

Figure 11 Width of the draining film in Zone IV for (*a*) glass, (*i*) $Q$ = 0.8 L min$^{-1}$; (*ii*) $Q$ = 1.5 L min$^{-1}$; and (*b*) polypropylene, (*i*) $Q$ = 2.3 L min$^{-1}$, (*ii*) $Q$ = 2.1 L min$^{-1}$. The nozzle diameter is $d_N$ = 3 mm. Also plotted are the loci predicted by the Mertens and Revised models for the contact angle $\beta_{fit}$.

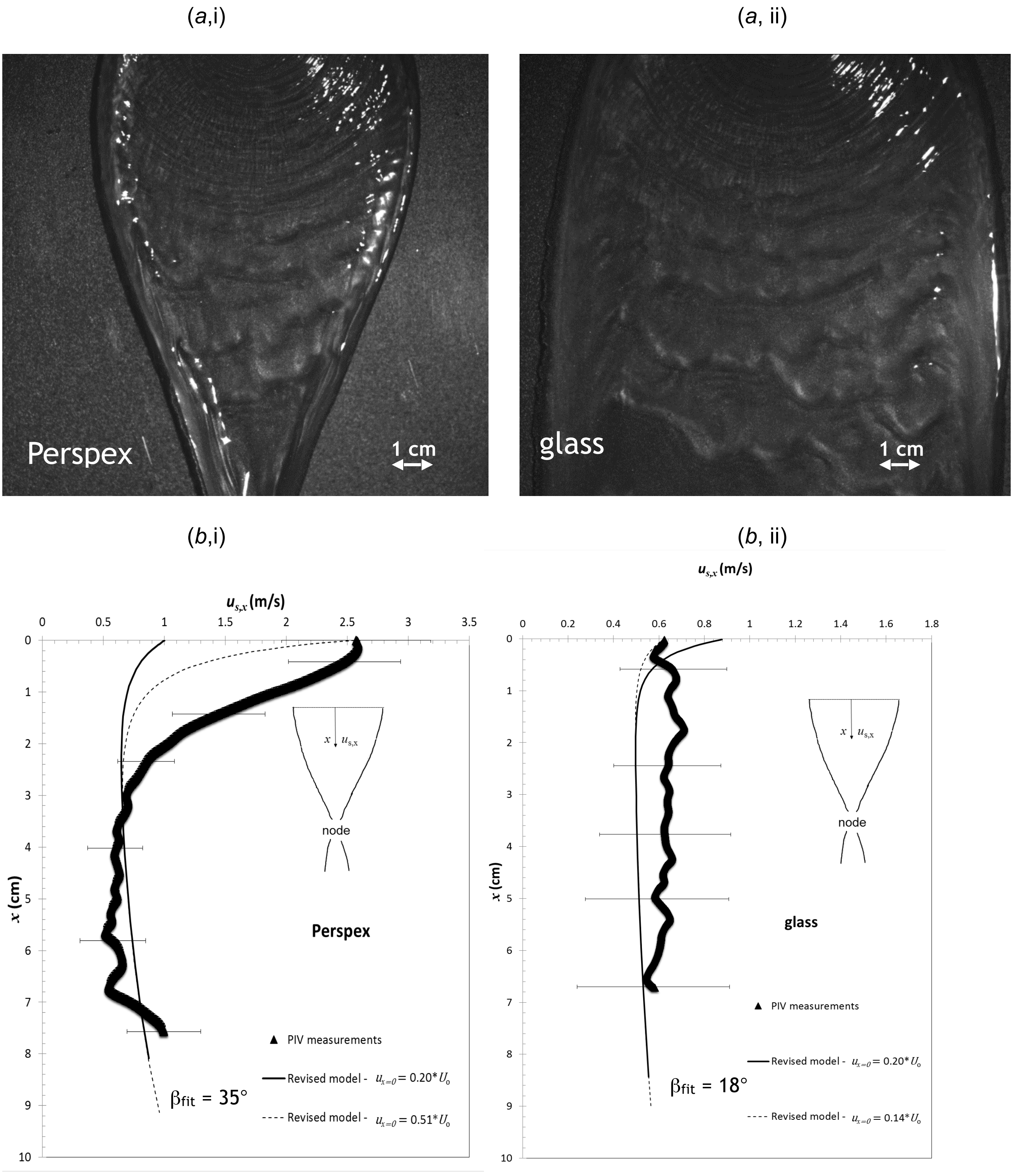

Figure 12 Comparison of zone IV behaviour: (*a*) photographs and (*b*) surface velocity, $u_s$, at centreline ($y$ = 0) for (*i*) Perspex, $Q$ = 0.63 L $min^{-1}$, and (*ii*) glass, $Q$ = 0.56 L $min^{-1}$. Plots in (*b*) show predictions for the Revised Model with different initial starting velocity, $u(x = 0)$ and effective contact angle, β, of (*i*) 35° and (*ii*) 18°.

(*a*)

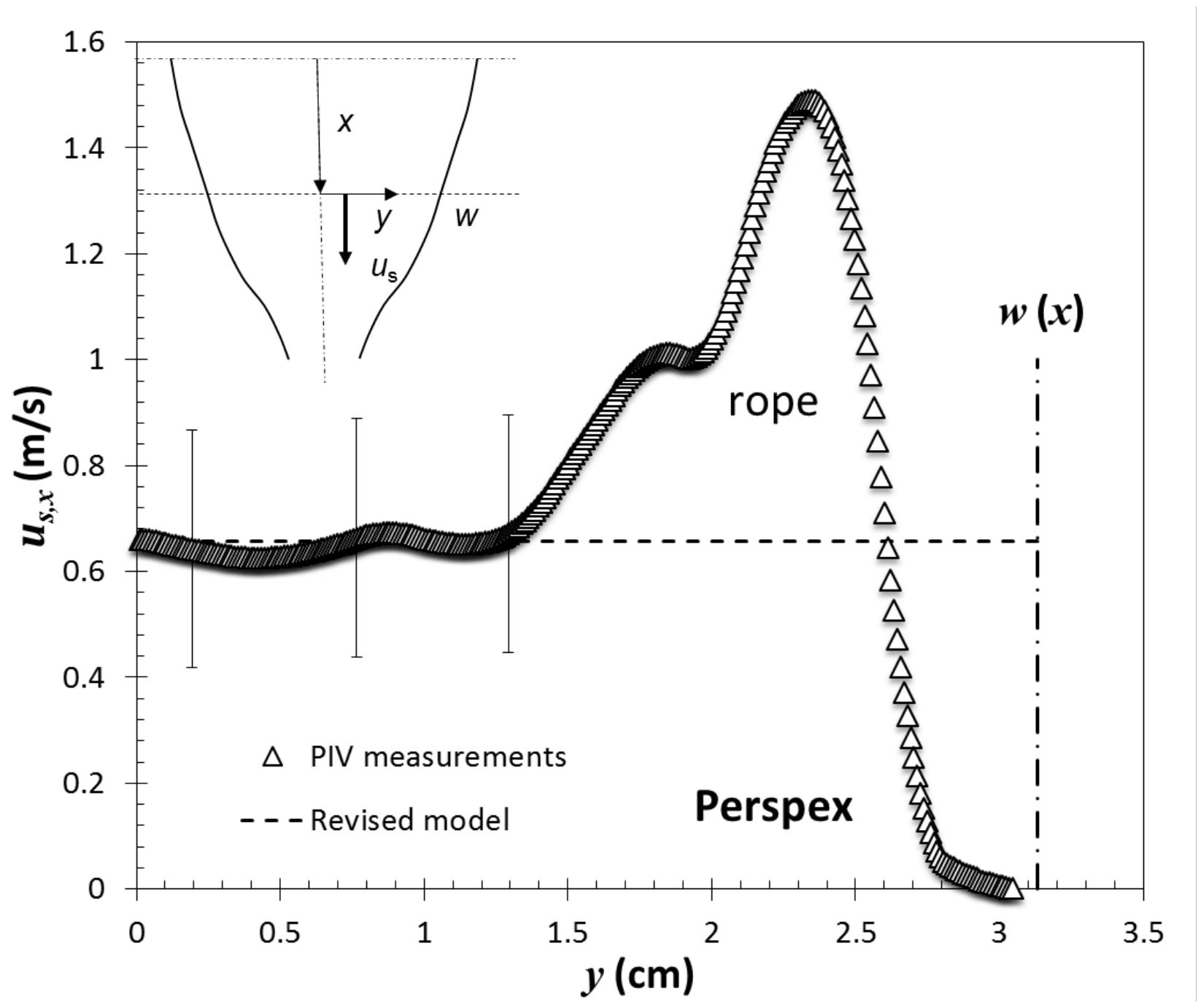

(*b*)

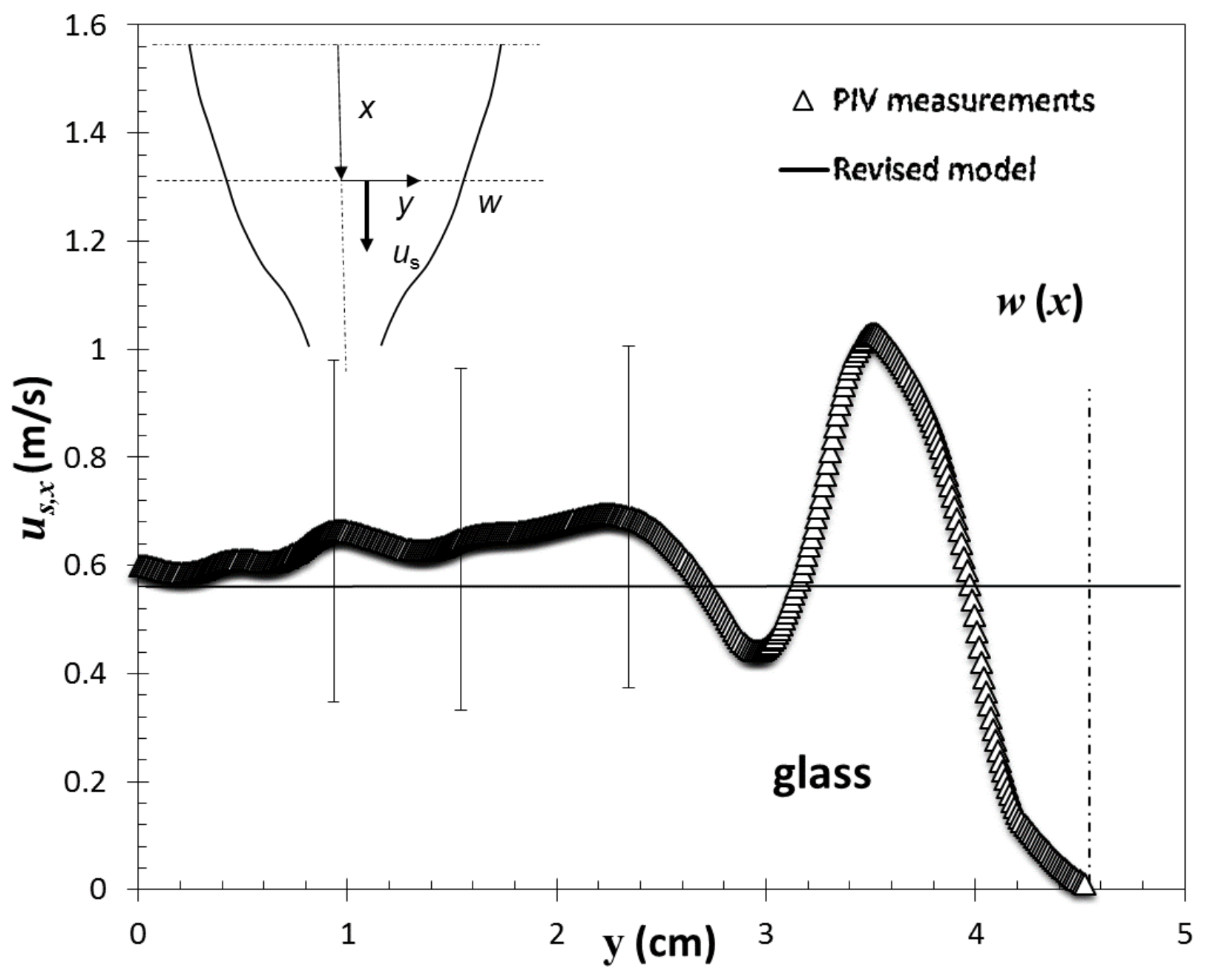

Figure 13 Comparison of measured downward surface velocity distribution, $u_s$, with value predicted by the Revised Model in Zone IV (horizontal locus) for (*a*) Perspex, $Q$ = 0.63 L min$^{-1}$, $x$ = 3.23 cm, β = 35°; (*b*) glass, $Q$ = 0.56 L min$^{-1}$, $x$ = 5.93 cm, .